\documentclass[iop]{emulateapj-rtx4}
\shortauthors{Sekanina \& Kracht}
\shorttitle{Comet C/1945 X1 (du Toit) --- a Dwarf Kreutz Sungrazer?}
\slugcomment{Version \today }
\newcommand{\Rsun}{$R_{\mbox{\scriptsize \boldmath $\odot$}}\!$}
\newcommand{\Rssun}{$R_{\mbox{\boldmath $\:\!\!\scriptstyle \odot$}}$}
\newcommand{\aster}{\mbox{$_{\displaystyle \ast}$}}

\newcommand{\gapeq}{$\;$\raisebox{0.3ex}{$>$}\hspace{-0.28cm}\raisebox{-0.75ex}{$\sim$}$\;$}

\begin{document}
\title{Was Comet C/1945 X1 (du Toit) a Dwarf, {\sl SOHO\/}-Like Kreutz
Sungrazer?}
\author{Zdenek Sekanina$^1$ \& Rainer Kracht$^2$}
\affil{$^1$Jet Propulsion Laboratory, California Institute of Technology,
  4800 Oak Grove Drive, Pasadena, CA 91109, U.S.A.\\
$^2$Ostlandring 53, D-25335 Elmshorn, Schleswig-Holstein, Germany}
\email{Zdenek.Sekanina@jpl.nasa.gov\\
{\hspace*{2.59cm}}R.Kracht@t-online.de{\vspace{-0.2cm}}}

\begin{abstract}
The goal of this investigation is to reinterpret and upgrade the astrometric
and other data on comet C/1945 X1, the least prominent among the Kreutz system
sungrazers discovered from the ground in the 20th century.  The central issue
is to appraise the pros and cons of a possibility that this object is ---
despite its brightness reported at discovery --- a dwarf Kreutz sungrazer.  We
confirm Marsden's (1989) conclusion that C/1945~X1 has a common parent with
C/1882~R1 and C/1965~S1, in line with the Sekanina-Chodas (2004) scenario of
their origin in the framework of the Kreutz system's evolution.  We integrate
the orbit of C/1882 R1 back to the early 12th century and then forward to
around 1945 to determine the nominal direction of the line of apsides and
perform a Fourier analysis to get insight into effects of the indirect
planetary perturbations.  To better understand the nature of C/1945 X1, its
orbital motion, fate, and role in the hierarchy of the Kreutz system, as well
as to attempt detecting the comet's possible terminal outburst shortly after
perihelion and answer the question in the title of this investigation, we
closely examined the relevant Boyden Observatory logbooks and identified both
the photographs with the comet's known images and nearly 20 additional patrol
plates, taken both before and after perihelion, on which the comet or traces
of its debris will be searched for, once the process of their digitization,
currently conducted as part of the Harvard College Observatory's DASCH Project,
has been completed and the scanned copies made available to the scientific
community.
\end{abstract}

\keywords{comets: general --- methods: data analysis{\vspace{0.1cm}}}

\section{Introduction}
As the most extensive system of genetically related comets in existence,
the Kreutz sungrazers represent an inexhaustible source of research
opportunities.  Kreutz's (1888, 1891, 1901) celebrated orbital studies
described the motions of the early bright members, showing that they moved
about the Sun in similar, extraordinarily elongated paths, with orbital
periods of up to about 1000 yr, yet approaching the Sun's surface to within
one solar radius at perihelion.  Kreutz's work was followed by many more
studies in the 20th century, with those by Marsden (1967, 1989) standing
out as the most important.

Because all Kreutz sungrazers are fragments of one progenitor, their orbits'
lines of apsides are nearly perfectly aligned; the scatter is only
a small fraction of 1$^\circ$, a product of indirect perturbations by
the planets, Jupiter in particular (Marsden 1967), and of the process of
cascading fragmentation (Sekanina 2002).  The research on the Kreutz system
has recently accelerated explosively thanks to vast new evidence provided by
imagers on board the spacecraft dedicated to the exploration of the Sun,
especially the coronagraphs C2 and C3 of the {\it Solar and Heliospheric
Observatory\/} ({\it SOHO\/}; see Brueckner et al.\ 1995) and the coronagraphs
COR2 and imagers HI1 of the {\it Solar Terrestrial Relations Observatory\/}'s
two probes ({\it STEREO-A\/} and {\it B\/}; see Howard et al.\ 2008).  Over the
past two decades, these instruments allowed detection, in close proximity of
the Sun, of thousands of minor Kreutz system's members, referred to hereafter
as the {\it dwarf\/} sungrazers, which keep streaming toward the Sun but
always disintegrate shortly before reaching perihelion.

The directions of the lines of apsides of 1600 dwarf Kreutz sungrazers derived
from their published gravitational orbits were recently shown by us (Sekanina
\& Kracht 2015; hereafter referred to as Paper 1) to be~distributed along an
arc of 25$^\circ$ (sic!) in the ecliptical latitude, failing utterly to comply
with the condition of directional alignment.  We determined that this major
effect was due to a neglected nongravitational acceleration in the dwarf
sungrazers' motions, which was orders of magnitude greater than the
nongravitational accelerations in the motions of the cataloged comets in
nearly-parabolic orbits with perihelia a few tenths of AU from the Sun or
more, topping in exceptional cases the Sun's gravitational acceleration.
In summary, the {\it preperihelion disintegration\/} and a {\it very high
erosion-driven nongravitational acceleration\/} of the orbital motion are
two fundamental attributes of the dwarf Kreutz sungrazers.

\section{Sungrazing Comet C/1945 X1 (du Toit)}
On 1945 December 11, a comet was discovered photo\-graphically by D.\ du Toit
at the Harvard College Observatory's Boyden Station near Bloemfontein, South
Africa (Paraskevopoulos 1945); nowadays this comet is referred to as C/1945
X1.  The object moved rapidly toward the Sun and its brightness at discovery
was reported as magnitude 7.  Additional plates were taken at Boyden on the
following nights until  December 15, but the five {\it estimated\/} astrometric
positions were not communicated until 1946 January 2, when Cunningham (1946a)
used them to compute three very preliminary parabolic orbits.  They indicated
that the comet was apparently a Kreutz sungrazer that had passed perihelion
five days before the cable was sent.  Cunningham (1946b) pointed out that his
search ephemeris, based on one of the three orbits, was uncertain ``by many
degrees.''  Published accounts show that after December 15 the comet was lost
and never seen again.  Contrary to expectations, it did not become a brilliant
object near and/or after perihelion.  Its failure to develop a bright headless
tail for a limited period of time after perihelion --- contrary to such
sungrazers as C/1887~B1 (see the references in Sekanina 2002) and C/2011 W3
(e.g., Sekanina \& Chodas 2012) --- suggests that C/1945 X1 may have
disintegrated already before perihelion, as do the dwarf Kreutz sungrazers.
This possibility raises a question of whether or not this object was the
only dwarf Kreutz sungrazer discovered and repeatedly observed
%
%
from the ground outside a total solar eclipse.\footnote{A dwarf sungrazer
{\vspace{-0.03cm}}C/2008 O1 was detected during a total solar eclipse on some
exposures after a search based on data from {\it SOHO\/} (Pasachoff et al.\
2009).  C/1882 K1 (Tewfik),~discovered $\sim$3.5~hr preperihelion at
5.6\,{\Rssun} (Marsden 1967, 1989) during a total solar eclipse on 1882 May 17,
may have been a dwarf sungrazer as well.} To address this issue in detail
requires that three critical points be answered:

(1) Can a very high nongravitational acceleration be conclusively detected in
the comet's orbital motion from the five Boyden observations?

(2) Why was the comet so bright 17 days before perihelion?  Was it in
outburst or was the reported brightness grossly overestimated?

(3) Was the absence of the comet's relics after perihelion conditioned on
a compelling qualifier or constraint, so that the comet's preperihelion
disintegration could be subject to doubt?

We present new evidence in the following sections that allows us to comment
on these points and to chart the lines of attack in the near future.

\section{The Boyden Photographic Observations}
A striking feature of the information on the comet's Boyden observations is
an extremely slow and protracted progress in propagating their results, with
an essentially complete lack of details.  This is most surprising, given that
C/1945~X1 was the first Kreutz sungrazer of the 20th century, after a pause
of nearly 60 years.

An emphatic example of the slow progress is the publication of the comet's
astrometry from the Boyden plates.  After a delayed message of the estimated
positions (whose observation times were announced with a precision to 1 hr!),
there was no follow-up report and no computation of an improved orbit until
22 years later, when Marsden (1967), revealing that the plates were measured
and reduced by A.\,G.\,Mowbray in 1952, derived several sets of parabolic
elements.  Strangely, these astrometric data were reported by neither Mowbray
himself nor L.\,E.\,Cunningham, for whom Mowbray was then working (Hockey
2009).  In fact, the positions were published only 37 years after they were
measured and reduced and 44 years after they were taken by the Boyden observers
(Marsden 1989)!

The circumstances of the comet's observations at Boyden have never been
published.  We were especially interested in the telescopes or cameras employed
to make these photographic observations, in exposure times used, and in the
type of tracking (sidereal or on the comet) of the post-discovery photographs.
The only information learnt from the literature was that in the 1940s Boyden
observers, such as M.\,J.\,Bester, discovered their comets with either the
Metcalf 25-cm f/4.9 Triplet refractor or the Bache 20-cm f/5.7 Doublet
refractor {\vspace{-0.02cm}}(Cooper 2003, 2005) while examining plates for
image quality.\footnote{These plates were taken on behalf of H.\,Shapley,
Director of the Harvard College Observatory, for the purpose of studying
southern variable stars (Cooper 2003, 2005; van Heerden 2008).}

As integral part of the Harvard College~Observatory's
\mbox{\it Digital\,Access\,to\,a\,Sky\,Century@Harvard\/}~(DASCH)~Proj\-ect,
all Boyden plates are in the process of being digitized, with the scans
gradually made available to the scientific community (Simcoe et al.\ 2006;
Grindlay et al.\ 2012).\footnote{The DASCH Project, at {\tt
http://dasch.rc.fas.harvard.edu/}, involves digitization of more than
500\,000 Harvard plates and is currently in progress; see also {\tt
http://tdc-www.harvard.edu/plates/}.} The products of this effort
will eventually play a major role in our quest for information on the
comet.  Of imminent interest to us are the observing logbooks on the
Harvard website; after some search, we have already been able to identify
all five plates measured and reduced by Mowbray.  Three surprises surfaced:\
(i)~the comet was {\it not\/} discovered on a plate taken with the Metcalf
or Bache telescope; (ii)~a never-reported post-perihelion search was
attempted with two instruments on 1946 January 8, 11 days after perihelion
and the same day that Cunningham's (1946b) ephemeris was issued; and
(iii)~none of the follow-up observations was made by the discoverer.
\begin{table*}[ht]
\begin{center}
\vspace{0.13cm}
{\footnotesize {\bf Table 1}\\[0.08cm]
Boyden Plates with Reported Images or Possible Images of Comet
C/1945 X1.\\[0.08cm]
\begin{tabular}{c@{\hspace{0.6cm}}l@{\hspace{0.07cm}}c@{\hspace{0.07cm}}r@{\hspace{0.7cm}}c@{\hspace{0.45cm}}c@{\hspace{0.6cm}}c@{\hspace{0.4cm}}c@{\hspace{0.3cm}}c@{\hspace{0.3cm}}c@{\hspace{0.6cm}}c}
\hline\hline\\[-0.22cm]
  & & & & \multicolumn{2}{@{\hspace{-0.4cm}}c}{Plate center$^{\rm b}$}
  & \multicolumn{2}{@{\hspace{-0.05cm}}c}{Sidereal time} & Exposure
  & & \\[-0.04cm]
Plate    & \multicolumn{3}{@{\hspace{-0.35cm}}c}{UT time at}
         & \multicolumn{2}{@{\hspace{-0.4cm}}c}{\rule[0.6ex]{2.55cm}{0.4pt}}
         & \multicolumn{2}{@{\hspace{-0.05cm}}c}{\rule[0.6ex]{2.25cm}{0.4pt}}
         & time & Logbook & \\[-0.04cm]
number\rlap{$^{\rm a}$} & \multicolumn{3}{@{\hspace{-0.35cm}}c}{mid-exposure}
         & \,\,R.A. &  \,Dec. & \,\,start & \,\,stop & (min) & reference
               & Observer$^{\rm c}$\\[0.05cm]
\hline\\[-0.25cm]
AM\,25201\rlap{$^{\rm d}$} & 1945 & Dec. &  11.04709 
             & $\!\!\!\!$13$^{^{\rm h}}$36\rlap{$^{^{\rm m}}\!\!\!$.7}
             & $\;-$60$^\circ 31^\prime$ &  7$^{^{\rm h}}$26\rlap{$^{^{\rm m}}$}
             & 8$^{^{\rm h}}$56\rlap{$^{^{\rm m}}$} & 90 & am45b\_0158
             & du Toit \\
AM\,25206    &  &  & 12.04713  & 15 \,18.7 & $-$65 \,22 & 7 $\:$30 & 9 \,00
             & 90 & am45b\_0158 & Bester \\
MF\,34987    &  &  & 13.07210  &  16 \,19.0 & $-$63 \,03 & 8 $\:$40 & 9 $\:$10
             & 30 & mf25b\_0086 & Britz \\
AM\,25208\rlap{$^{\rm e}$} &  &  & 14.03336  &  16 \,40.9 & $-$70 \,12
             & 7 $\:$43 & 8 $\:$23 & 40 & am45b\_0158 & J(?) \\
MF\,34991    &  &  & 14.05898  &  16 \,42.9 & $-$61 \,12 & 8 $\:$25 & 8 $\:$55
             & 30 & mf25b\_0086 & J(?) \\
MF\,34994    & & & 15.06906 & 17 \,02.6 & $-$58 \,39 & 8 $\:$51
             & 9 $\:$06 & 15 & mf25b\_0088 & J(?) \\
MF\,35030    & 1946 & Jan. & 8.06517 & 17 \,17.5 & $-$47 \,19 & 10 $\:05\;\;$
             & 10 $\:50\;\;$ & 45 & mf25b\_0094 & Bester \\
RB\,14184    &  &  &  8.07729 & 17 \,17.5 & $-$47 \,19 & 10 $\:25\;\;$
             & 11 $\:05\;\;$ & 40 & rb10\_148 & Bester \\[0.05cm]
\hline\\[-0.29cm]
\multicolumn{11}{l}{\parbox{14.8cm}{$^{\rm a}$\,{\scriptsize \mbox{AM} plates
taken with the Cooke lens; \mbox{MF} plates with the Metcalf telescope;
and \mbox{RB} plate with the Ross-Fecker camera.}}}\\[-0.09cm]
\multicolumn{11}{l}{\parbox{14.8cm}{$^{\rm b}$\,{\scriptsize Equinox
J2000.}}}\\[-0.04cm]
\multicolumn{11}{l}{\parbox{14.8cm}{$^{\rm c}$\,{\scriptsize The identity
of the observer whose abbreviation in the logbooks was J could not be
determined; see also Table 12.}}}\\[-0.12cm]
\multicolumn{11}{l}{\parbox{14.8cm}{$^{\rm d}$\,{\scriptsize Discovery
 plate.}}}\\[-0.05cm]
\multicolumn{11}{l}{\parbox{14.8cm}{$^{\rm e}$\,{\scriptsize This plate
 appears to have never been astrometrically measured and reduced.}}}\\[0.03cm]
\end{tabular}}
\end{center}
\end{table*}

Table 1 presents the results of our search for the relevant Boyden
plates.  For each, the individual columns list: the plate number, the UT
time of mid-exposure; the approximate equatorial coordinates of the
plate center (converted to the equinox J2000); the local sidereal times
of the exposure's start and termination; the resulting exposure time;
the logbook reference; and the observer.  Three instruments were employed
to observe, or search for, the comet:\ the Metcalf Triplet (plates MF;
a plate scale of 167$^{\prime\prime}\!$.3 mm$^{-1}$) and two small cameras
mounted piggyback on the Metcalf, the Cooke 3.8-cm f/8.9 lens (plates AM,
including the discovery one; a plate scale of 610$^{\prime\prime}\!$.8
mm$^{-1}$) and the Ross-Fecker 7.5-cm f/7 camera (plates RB; a plate scale
of 395$^{\prime\prime}\!$.5 mm$^{-1}$).  The limiting magnitudes of the
three instruments are listed on the Harvard website as, respectively, 17,
13--14, and 15, all much fainter than the comet's reported magnitude at
discovery.

All plates employed for C/1945 X1 had apparently a blue sensitive emulsion
(class L; see the DASCH website footnote)  and all three instruments used
a plate size of 20.3 cm (in right ascension) by 25.4 cm (in
declination),\footnote{The website~{\tt http://tdc-www.harvard.edu/plates/mf/\/}
specifies that Metcalf plates have generally sizes 8 in by 10 in, 14 in
by 17 in, and, some early ones, 10 in by 12 in.} covering fields of
15$^\circ\!\!$.1 (Metcalf), 35$^\circ\!\!$.7 (Ross-Fecker), and
55$^\circ\!\!$.2 (Cooke) in angular extent along a diagonal.

\begin{figure}[b]
\vspace{-11.55cm}
\hspace{0.95cm}
\centerline{
\scalebox{0.83}{
\includegraphics{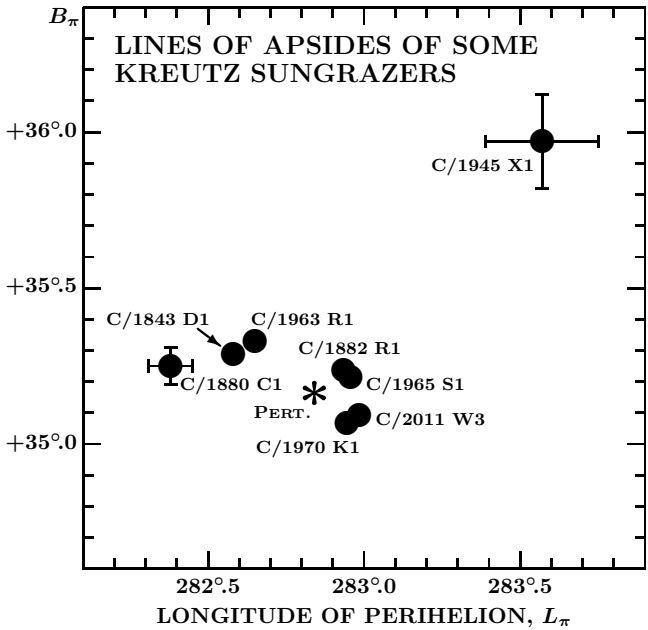}}} 
\vspace{-4.45cm}
\caption{Line of apsides for seven bright Kreutz sungrazers and C/1945 X1.
The plot of the perihelion longitude $L_\pi$ against perihelion latitude
$B_\pi$ shows that C/1945 X1, represented by Marsden's cataloged gravitational
orbit, deviates significantly from the cluster of the bright sungrazers,
whose lines of apsides are closely aligned.  The asterisk marked {\sc
Pert.}\ is the apsidal-line position for C/1945~X1 on the assumption that
the comet is a fragment of a common parent with C/1882~R1, the difference
being due entirely to the indirect planetary perturbations. For six comets the
errors of the coordinates $L_\pi$, $B_\pi$ are smaller than the size of
the symbols; the estimated errors for C/1880~C1 and C/1945~X1 are
depicted.{\vspace{0.22cm}}}
\end{figure}

The logbooks show that, of the five observations astrometrically
measured and reduced by Mowbray and published by Marsden (1989), the
ones on December 11--12 were made with the Cooke lens and only the ones
on December 13--15 with the Metcalf refractor.  The logbooks further
show that the comet was $\sim$12$^\circ$ from the center on the discovery
plate, 2$^\circ\!$.9 on the December 12 Cooke plate, always less than
1$^\circ\!$.2 on the Metcalf plates, and just about 9$^\circ\!$.6 on the
December 14 Cooke plate, which apparently has never been measured.  The
question of whether or not the comet is located on the plates taken on
January 8 can only be answered after our orbital analysis (see Sections 7.1
and 8.2).

\section{History of the Comet's Orbit Determination}
The sets of orbital elements for comet C/1945 X1 computed by Cunningham
(1946a) were much too uncertain to use as a basis of subsequent research.
A more accurate orbit was a gravitational parabolic approximation derived by
Marsden (1967), who employed the astrometry by Mowbray.  Marsden concluded
that the December 14 data point (which we found was measured on a MF plate)
was inconsistent with the other four positions and he omitted it from what
he described as the final set of orbital elements obtained by least
squares.  The four employed positions were fitted with a mean residual
of $\pm$2$^{\prime\prime}\!$.1, the orbit appeared to be similar to those
of C/1882 R1 and C/1965~S1, and ever since 1972 it has been listed, as the
representative orbit for C/1945~X1, in the {\it Catalogue of Cometary
Orbits\/} (see Marsden \& Williams 2008 for the most recent edition).

More recently, this comet's orbital motion was further examined by Marsden
(1989).  He computed four different gravitational solutions, A--D,
constraining them to a prescribed orientation of the line of apsides
and allowing the orbit to depart from a parabola.  His solutions C and D
are discussed more extensively in Section 6.1.\footnote{Marsden (1989)
remarked that the December 14 observation was reconstructed for an arbitrarily
chosen observation time from the residuals in his previous paper (Marsden
1967), because the original data were unfortunately lost.}

In terms of the orientation of the line of apsides, a very stable orbital
parameter, the representative orbit of C/1945 X1 is compared in Figure~1
with more reliably determined orbits of seven bright sungrazers.  The errors
for six of them are smaller than the size of the symbols.  For C/1945~X1 the
error is estimated by comparing the scatter in the apsidal line with that in
the angular elements, combined with the errors of the elements, as published
by Marsden (1967).  The estimated error for C/1880~C1 is due largely to the
uncertainty in the orbital period, as investigated by Kreutz (1901).  There
is a striking discrepancy between C/1945~X1 and the other sungrazers in
Figure~1, exceeding six standard deviations, thus providing grounds for
suspecting that the motion of C/1945~X1 might be --- like the motions of the
dwarf sungrazers --- significantly affected by nongravitational forces.

As of now, it has not been demonstrated conclusively whether there exists a
purely gravitational orbital solution for C/1945~X1 that simultaneously
(i)~fits satisfactorily at least four of the five available observations;
(ii)~complies with the proper orientation of the line of apsides; and
(iii)~is consistent with a plausible osculating orbital period.  Addressing
this issue requires in the first place that we know the appropriate values
of the quantities under (ii) and (iii), which depend on the perturbations
by the planets, Jupiter in particular.  We examine these perturbation
effects next.

\section{Indirect Perturbations by the Planets}
Because of the nature of their orbits, the Kreutz comets cannot
experience a close approach to the planets, including Jupiter. 
Nevertheless, the sungrazers' orbits are subjected to indirect planetary
perturbations over the entire revolution about the Sun and, as a result,
show limited variations.

Addressing this problem in some detail, Marsden (1967) began his
investigation of the effects of indirect perturbations on a fragment,
separating from its sungrazing parent at perihelion, by integrating the
fragment's motion over the orbital period to the time $t_\pi$ of next 
perihelion. First, he considered only Jupiter in a circular orbit in
the plane of the ecliptic.  He applied the equations for the variation
of arbitrary constants and found that in this simplified scenario the
fragment's orbital elements --- the argument of perihelion $\omega$,
the longitude of the ascending node $\Omega$, the inclination $i$, and
the perihelion distance $q$ --- at time $t_\pi$ depend on Jupiter's
ecliptical longitude at this same time, $\Lambda_{\rm J}(t_\pi)$, as
follows:
\begin{eqnarray}
\omega(\Lambda_{\rm J}) & = & \omega_0 +  X_\omega \sin (\Lambda_{\rm J}
  \!+\! \Lambda_0), \nonumber\\
\Omega(\Lambda_{\rm J}) & = & \Omega_0 + X_\Omega \sin (\Lambda_{\rm J}
  \!+\! \Lambda_0), \nonumber\\
i(\Lambda_{\rm J}) & = & i_0 + X_i \sin (\Lambda_{\rm J} \!+\! \Lambda_0),
 \nonumber\\
q(\Lambda_{\rm J}) & = & q_0 - X_q \sin (\Lambda_{\rm J} \!+\! \Lambda_0
  \!-\! 90^\circ) \nonumber \\[-0.05cm]
                  & = & q_0 + X_q \cos (\Lambda_{\rm J} \!+\! \Lambda_0),
\end{eqnarray}
where $\omega_0$, \ldots,\,$q_0$ are constant terms, $X_\omega$,
\ldots,\,$X_q$ are amplitudes (all taken as positive numbers), and $\Lambda_0$
is a constant phase~shift.  Because of Jupiter's nonzero orbital eccentricity
and the deviation of its orbital plane from the plane of the ecliptic, and also
because of the indirect perturbations by the other planets, the quasi-periodic
variations of the sungrazers' orbital elements are more complex.  Marsden
indicated that when only Jupiter's perturbations were accounted for, the
amplitudes amounted to approximately \mbox{$X_\omega = 1^\circ\!.1$},
\mbox{$X_\Omega = 1^\circ\!.4$}, \mbox{$X_i = 0^\circ\!.3$}, and \mbox{$X_q
= 0.00039\;{\rm AU} = 0.084\;${\Rsun}}. Marsden further pointed out that his
numerical integrations, which took into account the perturbations by the
planets Jupiter to Neptune and by Pluto, resulted in expressions similar to
Equation (1), with the amplitudes equal, respectively, to 1$^\circ\!$.6,
2$^\circ\!$.1, 0$^\circ\!$.4, and \mbox{0.00046 AU (= 0.099 {\Rsun})}.

Since the longitude and latitude of perihelion, $L_\pi$ and $B_\pi$, are
related to the three angular elements by
\begin{eqnarray}
\tan (L_\pi \!-\! \Omega) & = & \tan \omega \cos i, \nonumber\\
\sin B_\pi & = & \sin \omega \sin i,
\end{eqnarray}
the variation of the line of apsides is described, to a first approximation,
by:
\begin{eqnarray}
L_\pi(\Lambda_{\rm J}) & = & L_0 + X_L \sin (\Lambda_{\rm J} \!+\!
 \Lambda_0), \nonumber \\
B_\pi(\Lambda_{\rm J}) & = & B_0 + X_B \sin (\Lambda_{\rm J} \!+\!
 \Lambda_0),
\end{eqnarray}
where
\begin{eqnarray}
X_L & = & X_\Omega + (X_\omega \cos i_0 - X_i \cos \omega_0 \sin
B_0) \sec^2 \! B_0, \nonumber\\
X_B & = & (X_\omega \cot \omega_0 + X_i \cot i_0) \tan B_0 .
\end{eqnarray}

Marsden noticed that the line of apsides is scarcely{\pagebreak} affected by
the indirect perturbations.  To get a more profound insight into their
influence on the orbital motion of C/1945~X1, we first integrated the orbit
of its major presumed sibling, C/1882~R1{\vspace{-0.03cm}} (see Table~1
of Sekanina \& Chodas 2004),\footnote{This presumption is based both on
Marsden's (1967, 1989) and Sekanina \& Chodas' (2004) conclusions and on the
results of our own experimentation with the orbit of C/1945~X1, which showed
that it was much easier to align its apsidal orientation with that of C/1882~R1
than C/1843~D1, the other potential major sibling.{\vspace{-0.3cm}}} back in
time to the 1106 perihelion.  Next we varied the eccentricity incrementally and
ran a set of orbits, obtained in this fashion, forward in time.  Using the JPL
DE405 ephemeris, we accounted for the perturbations by all eight planets, by
Pluto, and by the three most massive asteroids, as well as for the relativistic
effect.  The increments were adjusted stepwise so that the perihelion times
ranged between 1939 August 2 and 1951 August~29 at nearly constant intervals
of about 100~days each; the length of the covered time period slightly exceeded
Jupiter's orbital period.  Checks showed that the relative errors accumulated
over an integration period of $\sim$1600~yr did not exceed $\sim$10$^{-10}$
in the comet's position vector and its velocity vector.

The integrated perturbations of the six orbital parameters, $\omega$, $\Omega$,
$i$, $L_\pi$, $B_\pi$, and $q$, for this period of sungrazers' arrival to
perihelion are presented in Figure 2 as a function of Jupiter's longitude
$\Lambda_{\rm J}$ at the perihelion times, $t_\pi$.  The variations are not
periodic in that the values of the elements at \mbox{$\Lambda_{\rm J} =
0^\circ$} in 1939 and 1951 differ by 0$^\circ\!$.70 in $\omega$, by
0$^\circ\!$.59 in $\Omega$, by 0$^\circ\!$.12 in $i$, by 0.068 {\Rsun} in
$q$, by 0$^\circ\!$.044 in $L_\pi$, and by 0$^\circ\!$.012 in $B_\pi$.  From
Figure 2 it follows that the variations for the last four parameters are
reasonably well approximated by the thick sine curve, while those for
$\omega$ and $\Omega$ are seen to deviate quite significantly, especially in
the range of Jupiter's longitudes from 30$^\circ$ to 150$^\circ$.  For at
least these two elements the scenario based on Jupiter in a circular orbit
in the plane of the ecliptic is clearly inadequate.

The perturbation variations in the perihelion longitude $L_\pi$ are a problem.
Equation (3) suggests that the curve should be in phase with the variations
in the argument of perihelion, the longitude of the ascending node, and the
inclination, but Figure 2 demonstrates that it is not.  Closer inspection shows
that the reason for this discrepancy is the fact that the predicted amplitude
given by Equation (3) is very close to zero (or, in fact, slightly negative)
and that the actual variations in $L_\pi$ in Figure~2 are determined by the
second order terms that have been neglected in Equation (3).  For the
perihelion latitude $B_\pi$ this is not the case and its variations are
indeed in phase with $\omega$, $\Omega$, and $i$.

The numerical integration shows that the net amplitudes of the orbital
parameters amount to 1$^\circ\!$.52 in $\omega$, 1$^\circ\!$.86 in $\Omega$,
0$^\circ\!$.39 in $i$, 0$^\circ\!$.11 in $L_\pi$ (which at a latitude of
$\sim$35$^\circ$ represents an arc of 0$^\circ\!$.09), 0$^\circ\!$.07 in
$B_\pi$, and 0.00051 AU or 0.110 {\Rsun} in $q$.  These are typically within
about 10\% of the amplitudes given by Marsden for the four elements from
Equation (1) above.  The overall amplitude for the line of apsides is
0$^\circ\!$.11, so that the range, 0$^\circ\!$.22, is generally consistent
with the apsidal difference of 0$^\circ\!$.30 between C/1882~R1 and C/1843~D1
accumulated over a span of two orbital periods (or about 15--17 centuries),
suggested for the age of the Kreutz system by Sekanina \& Chodas (2004, 2007).

\begin{figure*}[ht]
\vspace{-0.99cm}
\hspace{-0.64cm}
\centerline{
\scalebox{0.81}{
\includegraphics{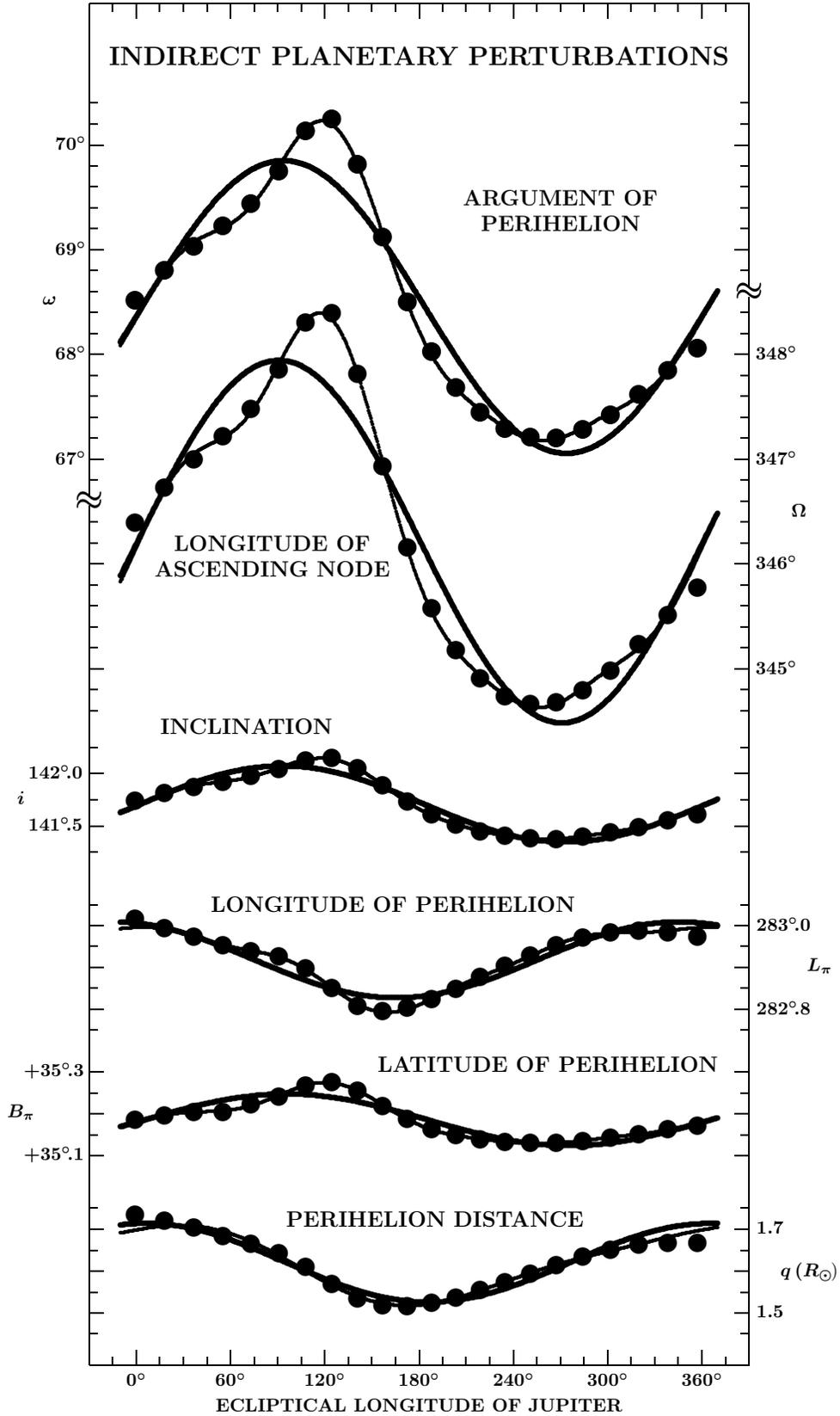}}} 
\vspace{-1.15cm}
\caption{Plots, against the ecliptical longitude of Jupiter $\Lambda_{\rm
J}$ at the time of a respective sungrazer's perihelion, of the argument of
perihelion $\omega$, the longitude of the ascending node $\Omega$, the
inclination $i$, the perihelion longitude $L_\pi$ and latitude $B_\pi$ (all
equinox J2000), and the perihelion distance $q$ of the siblings of comet
C/1882 R1 separated at the 1106 perihelion, which returned to perihelion
between 1939 August 2 and 1951 May 18.  The sungrazer C/1945 X1 was at
\mbox{$\Lambda_{\rm J}=195^\circ\!.23$}.  The thick sine curves show the
simple approximations of the type expressed by Equation (1), the thinner curves
are the much better approximations by the $N$-harmonic Fourier polynomials
(\mbox{$N = 4$} for all but the last curve, for which \mbox{$N = 2$}; see
Table 3).  Note that the scales for $L_\pi$ and $B_\pi$ are four times wider
than those for $\omega$, $\Omega$, and $i$.}
\end{figure*}

We now constrain the orbital properties of C/1945 X1 by its arrival time at
perihelion, 1945 December 28.0 TT.  The predicted orientation of
the comet's line of apsides is determined by the ecliptical coordinates
\mbox{$L_\pi = 282^\circ\!.84$} and \mbox{$B_\pi = +35^\circ\!.16$} (Equinox
J2000), deviating from the orientation of the apsidal line of C/1882 R1
by 0$^\circ\!$.11.  The longitude of Jupiter at the time of the comet's
perihelion is \mbox{$\Lambda_{\rm J} = 195^\circ\!.23$}.  In addition,
the computations show that the comet's osculating semimajor axis should
have been 96.7 AU and the orbital period 951 yr, longer than the actual
time span since 1106.  This orbital period is in excellent agreement with
the value that Sekanina \& Chodas (2004) obtained for this comet in their
hierarchy model of the Kreutz system (see their Table 12).

\begin{table}[t]
\begin{center}
\vspace{0.1cm}
{\footnotesize {\bf Table 2}\\[0.1cm]
Cataloged Orbit of C/1945 X1 and Orbit of the Hybrid\\(Equinox J2000).\\[0.15cm]
\begin{tabular}{l@{\hspace{0.4cm}}c@{\hspace{0.4cm}}c}
\hline\hline\\[-0.22cm]
                & C/1945 X1             & Hybrid$^{\rm b}$ \\[-0.05cm]
Orbital element & (cataloged$^{\rm a}$) & (perturbed) \\[0.05cm]
\hline\\[-0.17cm]
Perihelion time $t_\pi$ (1945 TT) & Dec.\,27.9652 & Dec.\,27.9801 \\
Argument of perihelion $\omega$ & $\;\:$72$^\circ\!$.0619
 & $\;\:$67$^\circ\!$.8518 \\
Longitude of ascending node $\Omega$ & 351$^\circ\!$.2006
 & 345$^\circ\!$.3759 \\
Orbit inclination $i$ & 141$^\circ\!$.8734 & 141$^\circ\!$.5608 \\
\hspace{3.05cm}(AU) & 0.007516 & 0.007123 \\[-0.33cm]
Perihelion distance $q \left\{ \raisebox{0ex}[0.3cm][0.3cm]{}
 \right.$\\[-0.39cm]
\hspace{3.05cm}(\Rssun) & 1.6147 & 1.5302 \\
Orbit eccentricity & 1.0 & 0.99992634 \\
Semimajor axis (AU) & {\scriptsize $\infty$} & 96.7 \\
Orbital period (yr) & {\scriptsize $\infty$} & 951 \\
Longitude of perihelion $L_\pi$ & $\;$283$^\circ\!$.57
 & $\;$282$^\circ\!$.84 \\
Latitude of perihelion $B_\pi$ & +35$^\circ\!$.97 & +35$^\circ\!$.16 \\
Osculation epoch (1945 TT) &   (none)   &  Dec.\ 26.0 \\[0.05cm]
\hline\\[-0.24cm]
\multicolumn{3}{l}{\parbox{8cm}{$^{\rm a}$\,{\scriptsize See Marsden \&
 Williams (2008).}}} \\[-0.1cm]
\multicolumn{3}{l}{\parbox{8cm}{$^{\rm b\!}$\,{\scriptsize Orbit of
 C/1882 R1 integrated to perihelion time of C/1945 X1.}}} \\[0.12cm]
%
\end{tabular}}
\end{center}
\end{table}

To illustrate the influence of the indirect perturbations specifically on
the orbit of C/1945 X1, we compare Marsden's (1967) cataloged parabolic orbit
(cf.\ Marsden \& Williams' 2008) with the orbit of what we call a {\it
hybrid\/} ---  one that C/1882 R1 would have had, if it arrived at perihelion
at the time C/1945 X1 did.  The two orbits differ by several degrees in the
argument of perihelion and the longitude of the ascending node, and by almost
0.1 {\Rsun} in the perihelion distance.

Returning now to Figure 2, one could think of the differences between the
sequence of the points, computed by numerical integration of the orbits,
and the thick sine curves that are intended to fit them, as deformations and,
accordingly, employ the superposition principle to mitigate the discrepancies.
The effects of the perturbations, even though {\it not\/} periodic on a scale
of Jupiter's orbital period, can nonetheless be expressed as a combination
of periodic variations in terms of an $N$-harmonic Fourier polynomial.
Figure 2 shows the results of this fitting, with $N$ not exceeding 4, as the
lighter curves; obviously, the improvement over a simple sine curve is
considerable.  Thus, calling $\Re$ any of the six parameters in Equations (1)
and (3), we can write it in the form:
\begin{eqnarray}
\Re(\Lambda_{\rm J}) & = & \Re_0 + \sum_{k=1}^{N} \,\, [a_{\Re,k} \sin (k
  \Lambda_{\rm J}) + b_{\Re,k} \cos (k \Lambda_{\rm J})] \nonumber \\[-0.1cm]
                     & = & \Re_0 + \sum_{k=1}^{N} X_{\Re,k} \sin (k
  \Lambda_{\rm J} \!+\! \Lambda_{\Re,k-1}),
\end{eqnarray}
where
\begin{eqnarray}
X_{\Re,k} & = & \sqrt{a_{\Re,k}^2 \!+ b_{\Re,k}^2}, \nonumber \\
\Lambda_{\Re,k-1} & = & \arctan \! \left( \! \frac{b_{\Re,k}}{a_{\Re,k}} \!
 \right).
\end{eqnarray}
It is apparent that for \mbox{$N = 1$}, Equation (5) emulates Equations (1)
and (3), with $\Lambda_{\Re,0}$ being always equal to $\Lambda_0$ for
\mbox{$\Re = (\omega, \Omega, i, L_\pi, B_\pi)$} and \mbox{$\Lambda_{\Re,0}
= \Lambda_0 \!+\!  90^\circ$} for \mbox{$\Re = q$}, and with \mbox{$X_{\Re,1}
= X_\Re$} for any orbital parameter.

The coefficients $a_{\Re,k}$ and $b_{\Re,k}$ are computed by least squares
from the values of $\Re(\Lambda_{\rm J})$ derived from the orbit integration
runs as a function of Jupiter's longitude $\Lambda_{\rm J}(t_\pi)$.  The
appropriate number of terms $N = N_{\rm min} > 1$ is determined by a minimum
mean residual.  As the number of Fourier coefficients in the equations of
condition increases with $N$, the number of the degrees of freedom drops,
thus causing the mean residual $\sigma$ to increase.  A byproduct is an
increasing uncertainty of the Fourier coefficients in the (meaningless)
polynomials with \mbox{$N > N_{\rm min}$}.  As examples, the plots of the
mean residuals for $\omega$, $\Omega$, and $i$ vs $N$ are displayed in
Figure 3.

\begin{figure}[t]
\vspace{-3.1cm}
\hspace{0.98cm}
\centerline{
\scalebox{0.85}{
\includegraphics{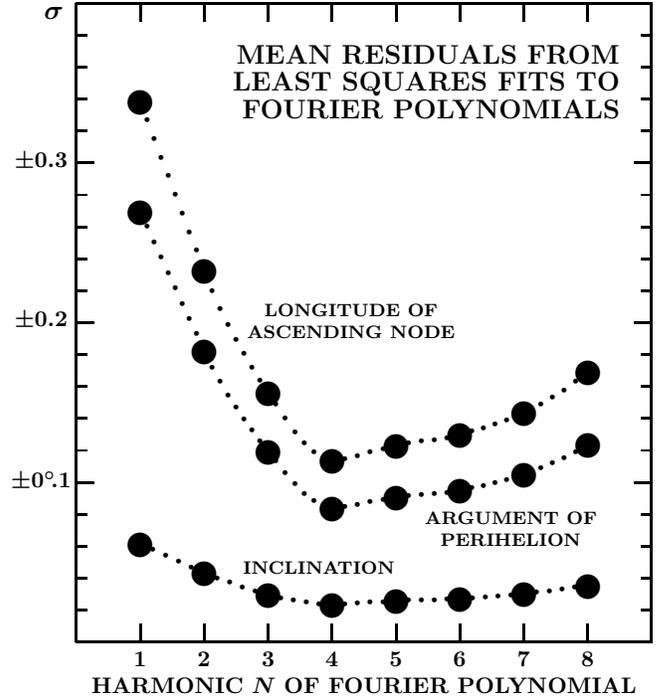}}} 
\vspace{-13.05cm}
\caption{Mean residuals $\sigma$ of the Fourier polynomials, fitted to the
variations in the longitude of the ascending node, the argument of perihelion,
and the inclination during Jupiter's orbital period 1939--1951, as a function
of the polynomials' harmonic $N$.  The minimum residual for all three orbital
elements is reached at \mbox{$N = 4$}.{\vspace{0.05cm}}}
\end{figure}

\begin{table*}[t]
\begin{center}
\vspace{0.1cm}
{\footnotesize {\bf Table 3}\\[0.05cm]
Fourier Polynomials Fitting the Dependence of Indirect Planetary Perturbations
 of Orbital Elements of\\Kreutz Sungrazers in 1939--1951 on Ecliptical
 Longitude of Jupiter.\\[0.1cm]
\begin{tabular}{l@{\hspace{0.5cm}}c@{\hspace{0.55cm}}c@{\hspace{0.55cm}}c@{\hspace{0.1cm}}c@{\hspace{0.1cm}}c@{\hspace{0.6cm}}c}
\hline\hline\\[-0.2cm]
& & Mean & Constant & Harmonic & Harmonic & Harmonic \\
Orbital element & Fit$^{\rm a}$ & residual & term$^{\rm b}$
& $k$\,=\,1,\,\ldots,$N$ & amplitude & phase \\[0.07cm]

\hline\\[-0.22cm]
Argument of perihelion, $\omega$ & A & $\pm$0$^\circ\!\!$.269 &
68$^\circ\!\!$.454 $\pm$ 0$^\circ\!\!$.057 & \,\ldots & 1$^\circ\!\!$.397
$\pm$ 0$^\circ\!\!$.083 & 356$^\circ\!\!$.2 $\pm$ 3$^\circ\!\!$.3 \\
& F$_4$ & $\pm$0.084 & 68.455 $\pm$ 0.018 & 1 & 1.386 $\pm$ 0.026 & 355.9
$\pm$ 1.0 \\
& & & & 2 & 0.272 $\pm$ 0.026 & 202.1 $\pm$ 5.3 \\
& & & & 3 & 0.183 $\pm$ 0.025 & $\;\:$57.3 $\pm$ 8.0 \\
& & & & 4 & 0.105 $\pm$ 0.025 & $\;\:$319 $\pm$ 14 \\[0.08cm]
Longitude of ascending node, $\Omega$ & A & $\pm$0.338 & 346.214 $\pm$
0.072 & \,\ldots & 1.731 $\pm$ 0.104 & 359.1 $\pm$ 3.3 \\
& F$_4$ & $\pm$0.113 & 346.216 $\pm$ 0.024 & 1 & 1.718 $\pm$ 0.035 & 358.8
$\pm$ 1.1 \\
& & & & 2 & 0.338 $\pm$ 0.035 & 205.8 $\pm$ 5.8 \\
& & & & 3 & 0.229 $\pm$ 0.034 & $\;\:$59.0 $\pm$ 8.7 \\
& &  & & 4 & 0.134 $\pm$ 0.035 & $\;\:$323 $\pm$ 15 \\[0.08cm]
Orbit inclination, $i$ & A & $\pm$0.061 & 141.709 $\pm$ 0.013 & \,\ldots &
0.364 $\pm$ 0.019 & 357.0 $\pm$ 2.8 \\
& F$_4$ & $\pm$0.023 & 141.710 $\pm$ 0.005 & 1 & 0.361 $\pm$ 0.007
& 356.7 $\pm$ 1.1 \\
& & & & 2 & 0.059 $\pm$ 0.007 & 198.6 $\pm$ 6.7 \\
& & & & 3 & 0.042 $\pm$ 0.007 & $\;\:$52.0 $\pm$ 9.6 \\
& & & & 4 & 0.024 $\pm$ 0.007 & $\;\:$320 $\pm$ 17 \\[0.08cm]
Longitude of perihelion, $L_\pi$ & A & $\pm$0.0202 & 282.9179 $\pm$
0.0043 & \,\ldots & 0.0968 $\pm$ 0.0060 & 104.6 $\pm$ 3.9 \\
& F$_4$ & $\pm$0.0087 & 282.9183 $\pm$ 0.0019 & 1 & 0.0906 $\pm$ 0.0026 &
104.1 $\pm$ 1.7 \\
& & & & 2 & 0.0222 $\pm$ 0.0026 & 299.2 $\pm$ 6.9 \\
& & & & 3 & 0.0077 $\pm$ 0.0027 & $\;\:$143 $\pm$ 20 \\
& & & & 4 & 0.0082 $\pm$ 0.0027 & $\;\;\:\:$29 $\pm$ 18 \\[0.08cm]
Latitude of perihelion, $B_\pi$ & A & $\pm$0.0169 & +35.1860 $\pm$
0.0036 & \,\ldots & 0.0616 $\pm$ 0.0052 & 354.4 $\pm$ 4.6 \\
& F$_4$ & $\pm$0.0030 & +35.1859 $\pm$ 0.0006 & 1 & 0.0608 $\pm$ 0.0009 &
353.8 $\pm$ 0.9 \\
& & & & 2 & 0.0185 $\pm$ 0.0009 & 194.6 $\pm$ 2.7 \\
& & & & 3 & 0.0111 $\pm$ 0.0009 & $\;\:$66.3 $\pm$ 4.7 \\
 & & & & 4 & 0.0055 $\pm$ 0.0009 & 321.4 $\pm$ 9.3 \\[0.08cm]
Perihelion distance, $q$ (\Rssun) & A & $\pm$0.0156 & 1.6200 $\pm$
0.0034 & \,\ldots & 0.0987 $\pm$ 0.0048 & $\;\:$84.0 $\pm$ 2.9 \\
& F$_2$ & $\pm$0.0132 & 1.6183 $\pm$ 0.0028 & 1 & 0.0905 $\pm$ 0.0039 &
$\;\:$83.2 $\pm$ 2.6 \\
& & & & 2 & 0.0173 $\pm$ 0.0040 & $\;\:$325 $\pm$ 13 \\[0.05cm]
\hline \\[-0.2cm]
\multicolumn{7}{l}{\parbox{15.8cm}{$^{\rm a}$\,{\scriptsize A = approximation
by Equation (1) or (3), based on the assumption of a single{\vspace{-0.03cm}}
planet in circular orbit in the plane of the~ecliptic; F$\!_N$ =
$N$-harmonic Fourier polynomial that provides the best fit (minimum mean
residual).}}}\\[0cm]
\multicolumn{7}{l}{\parbox{15.8cm}{$^{\rm b}$\,{\scriptsize Equinox
J2000.}}}\\[0.1cm]
\end{tabular}}
\end{center}
\end{table*}

The Fourier solutions for the six orbital parameters in Figure 2 are listed
in Table 3.  For each parameter the first line provides the results of the
approximate solution, fit A, by applying Equation (1) or (3).  The second
and following lines refer to the fitted Fourier polynomial, F$\!_N$.  The
mean residual $\sigma$ is always followed by the constant term $\Re_0$ and
by information on each harmonic $k$ in the case of the $N$-harmonic Fourier
polynomial:\ the harmonic amplitude $X_{\Re,k}$ and phase $\Lambda_{\Re,k-1}$.

We note that the amplitudes of all harmonics are determined with an error
not exceeding $\pm$35\%, but mostly less than $\pm$25\%.  With one borderline
exception the amplitudes decrease with increasing harmonic, as expected.
The amplitude of the first harmonic is always only slightly smaller than
the net amplitude derived directly from the results of numerical integration
of the orbit.  The phase angles of all harmonics for the argument of
perihelion, the longitude of the ascending node, the inclination, and the
latitude of perihelion are remarkably close to one another. The first harmonic
for the perihelion distance is shifted, as expected, by almost exactly
90$^\circ$ relative to those of $\omega$, $\Omega$, $i$, and $B_\pi$.  Only
the longitude of perihelion $L_\pi$ appears to be out of phase with any of
the other orbital parameters.

The least-squares procedure also allows us to generalize Equation (5) thus:
{\vspace{-0.2cm}}
\begin{equation}
\Re(\Lambda_{\rm J}) = \Re_0 + \sum_{k=1}^{N} X_{\Re,k} \sin (c_k
\Lambda_{\rm J} \!+\! \Lambda_{\Re,k-1}),
\end{equation}
where $c_k$'s are any positive numbers, not restricted to integers.  This
generalization offers us an opportunity to relate the coefficients $c_k$
($k > 1$) to the orbital periods of other planets, $P_k$, so that
\mbox{$c_k = P_{\rm J}/P_k$}.  Without providing any details, we point
out that our search for least-squares solutions of this type proved much
less successful in fitting the quasi-periodic variations in the sungrazers'
orbital elements than did the Fourier polynomials.

\begin{table*}[ht]
\begin{center}
\vspace{0.13cm}
{\footnotesize {\bf Table 4}\\[0.08cm]
Measured and Reduced Boyden Observations of Comet C/1945 X1 (Equinox
J2000).\\[0.1cm]
\begin{tabular}{c@{\hspace{0.6cm}}l@{\hspace{0.07cm}}r@{\hspace{0.6cm}}c@{\hspace{0.6cm}}c@{\hspace{0.5cm}}c@{\hspace{0.35cm}}c@{\hspace{0.65cm}}c@{\hspace{0.7cm}}c@{\hspace{0.45cm}}c}
\hline\hline\\[-0.22cm]
& & & & & & & & \multicolumn{2}{@{\hspace{0.02cm}}c}{From plate center} \\[-0.1cm]
Ref. & \multicolumn{2}{@{\hspace{-0.4cm}}c}{Observation time} & & & Assigned
 & Instrument & Plate & \multicolumn{2}{@{\hspace{0.02cm}}c}{\rule[0.3ex]{2.45cm}{0.4pt}} \\
No. & \multicolumn{2}{@{\hspace{-0.4cm}}c}{1945 (UT)} & R.A. & Dec. & weight
    & employed & number & \,distance & $\;$P.A. \\[0.08cm]
\hline\\[-0.27cm]
1 & December & 11.04687 & $\:\!$15$^{^{\rm h}}\:\!\!\!$13$^{^{\!\rm
  m}}\:\!\!\!$24$^{^{\rm s}}\:\!\!\!\!$.42
  & $\:\!-$65$^{\:\!\!^\circ}\!$23$^{^\prime}$43$^{^{\prime\:\!\!\prime}
  \:\!\!\!\!\!}$.4 & 1.0 & Cooke & AM\,25201 & $\!\!$11$^\circ\!\!$.94
  & \,124$^\circ\!\!$.4 \\
2 &          & 12.04691 & 15 43 21.25 & $-$64 09 13.1 & 1.0 & Cooke
  & AM\,25206 & 2.89 & $\;\:$68.0 \\
3 &          & 13.07189 & 16 11 04.21 & $-$62 29 32.5 & 3.6 & Metcalf
  & MF\,34987 & 1.07 & 300.5 \\
4 & & 14.07000\rlap{$^{\rm a}$} & 16 34 56.77 & $-$60 33 16.5 & 3.6 & Metcalf
  & MF\,34991 & 1.16 & 302.6 \\
5 &          & 15.06885 & 16 55 41.67 & $-$58 22 47.2 & 3.6 & Metcalf
  & MF\,34994 & 0.95 & 285.7 \\[0.05cm]
\hline\\[-0.2cm]
\multicolumn{10}{l}{\parbox{14cm}{$^{\rm a}$\,{\scriptsize Position
reconstructed by Marsden for an arbitrarily chosen time on this date (see
footnote in Section 4).{\vspace{0.1cm}}}}}
\end{tabular}}
\end{center}
\end{table*}

\section{Orbital Computations Based on the Boyden Observations from December
11--15}
If comet C/1945 X1 is closely related to C/1882~R1, the hybrid's orbit in
Table 3 suggests that the line of apsides be described by \mbox{$L_\pi =
282^\circ\!\!$.84} and \mbox{$B_\pi = +35^\circ\!\!$.16} (Equinox J2000),
as plotted in Figure~1 (marked {\sc Pert.}), and that the osculating
semimajor axis at perihelion be equal to \mbox{$a = 96.7$ AU}.  Table 3
also shows that the osculating elements should differ from those of
C/1882~R1 by $\sim$2$^\circ$ in $\omega$ and $\Omega$, by almost
0$^\circ\!$.5 in $i$, and by more~than 0.1 {\Rsun} in $q$.  Below we
investigate a range of orbital solutions, gravitational and nongravitational,
offered by the observations made at Boyden on December 11--15.

As measured and reduced by Mowbray and listed for the equinox B1950 by Marsden
(1989), the five astrometric positions are for the equinox J2000 presented in
Table~4.  Because on the first two nights the comet was imaged with the Cooke
lens, while on the three remaining nights with the Metcalf refractor, we assign
the positional data weights that are inversely proportional to the plate scales
of the two instruments.  Comparison of the observation times in column~2 of
Table 4 with those in column~2 of Table 1 shows that the first ones are
systematically too early by 0.00021 to 0.00022 day, or 18--19~s.  The nature
of the difference is unknown; it cannot come from a longitude discrepancy,
because it is equivalent to a distance of as much as $\sim$500 meters.  The
UT times in Table 1, determined from the sidereal times of the beginning and
end of the exposures, are in fact UT1 times and have independently been
checked; they never left an unexplained difference of more than $\sim$1~s.
Nevertheless, for the sake of comparison with Marsden's (1967, 1989) results,
we retain for this exercise his UT times.

\subsection{Gravitational Solutions}
Since the observed orbital arc is too short to reliably compute the
eccentricity, it needs to be determined from other considerations.  As
already mentioned above, the constraint is here provided by adopting the
semimajor axis of the hybrid's osculating orbit, 96.7 AU at the epoch of
1945 December 26.0 TT, for all solutions that follow.  We begin with a
gravitational orbit to fit all five astrometric positions in Table 4.

Deriving orbits B and C and comparing them with orbit A, Marsden (1989)
demonstrated that elliptical solutions fitted the five observations of
C/1945 X1 much better than a parabolic approximation.  Yet, neither orbit,
B or C, provides a satisfactory fit.  One possible contributor to this
problem could be the constraint introduced to satisfy the prescribed
direction of the line of apsides (for which Marsden took \mbox{$L_\pi =
282^\circ\!.7$}, \mbox{$B_\pi = +35^\circ\!.2$} after conversion to Equinox
J2000), the same issue that we encountered with the dwarf sungrazers of the
Kreutz system (Paper~1; see also Section 1).

\begin{table}[b]
\begin{center}
\vspace{0.4cm}
{\footnotesize {\bf Table 5}\\[0.08cm]
Marsden's (1989) Orbit C and Our Optimum Gravitational Solution to Fit All
Five Boyden Observations of\\C/1945 X1 (Equinox J2000).\\[0.08cm]
\begin{tabular}{l@{\hspace{0.4cm}}c@{\hspace{0.4cm}}c}
\hline\hline\\[-0.22cm]
Orbital element & Orbit C & {\it Best Fit} \\[0.1cm]
\hline\\[-0.2cm]
Perihelion time $t_\pi$ (1945 TT) & Dec.\,27.976 & Dec.\,27.982 \\
Argument of perihelion $\omega$ & $\;\:$68$^\circ\!$.10
 & $\;\:$70$^\circ\!$.99 \\
Longitude of ascending node $\Omega$ & 345$^\circ\!$.54
 & 349$^\circ\!$.50 \\
Orbit inclination $i$ & 141$^\circ\!$.60 & 141$^\circ\!$.91 \\
\hspace{3.05cm}(AU) & 0.00699 & 0.007244 \\[-0.33cm]
Perihelion distance $q \left\{ \raisebox{0ex}[0.3cm][0.3cm]{}
 \right.$\\[-0.39cm]
\hspace{3.05cm}(\Rssun) & 1.502 & 1.556 \\
Orbit eccentricity & 0.99993010 & 0.99992506 \\
Semimajor axis (AU) & 100 & 96.67 \\
Orbital period (yr) & 1000 & 950.5 \\
Longitude of perihelion $L_\pi$ & $\;$282$^\circ\!$.70
 & $\;$283$^\circ\!$.14 \\
Latitude of perihelion $B_\pi$ & +35$^\circ\!$.20 & +35$^\circ\!$.68 \\
Osculation epoch 1945 (TT) &   (none)   &  Dec.\ 11.0 \\
Mean residual$^{\rm a}$ & $\pm$11$^{\prime\prime}\!\!$.95
                        & $\pm$6$^{\prime\prime}\!\!$.17 \\[0.05cm]
\hline\\[-0.24cm]
\end{tabular}
\begin{tabular}{l@{\hspace{0.07cm}}r@{\hspace{1cm}}c@{\hspace{0.5cm}}c@{\hspace{0.8cm}}c@{\hspace{0.5cm}}c}
\multicolumn{6}{@{\hspace{0.1cm}}c}{Distribution of Residuals$^{\rm a}$,
 $O \!-\! C$} \\[0.1cm]
\hline \\[-0.24cm]
\multicolumn{2}{@{\hspace{-0.7cm}}c}{Time of}
& \multicolumn{2}{@{\hspace{-0.65cm}}c}{Orbit C}
& \multicolumn{2}{@{\hspace{-0.05cm}}c}{\it Best Fit} \\[-0.05cm]
\multicolumn{2}{@{\hspace{-0.7cm}}c}{observation}
& \multicolumn{2}{@{\hspace{-0.65cm}}c}{\rule[0.6ex]{2.15cm}{0.4pt}}
& \multicolumn{2}{@{\hspace{-0.05cm}}c}{\rule[0.6ex]{2.15cm}{0.4pt}} \\
\multicolumn{2}{@{\hspace{-0.7cm}}c}{1945 (UT)} & R.A. & Dec. & R.A.
  & Dec. \\[0.1cm]
\hline\\[-0.15cm]
Dec. & 11.04687 & \,\,+5$^{\prime\prime}\!\!$.6 & $\!$+22$^{\prime\prime}\!\!$.4
 & $\!$+13$^{\prime\prime}\!\!$.2 & $\;$+8$^{\prime\prime}\!\!$.0 \\ 
     & 12.04691 & +6.1 & \,+5.3 & +6.1 & +2.5 \\
     & 13.07189 & $\!\!\!-$14.8 & $\!\!-$11.8 & $-$6.6 & $-$5.6 \\
     & 14.07000 & $\!\!\!$+10.3 & \,$-$1.9 & +7.8 & +8.0 \\
     & 15.06885 & $\!\!\!$+15.6 & $\!\!-$10.9 & $-$2.7 & $-$3.2 \\[0.05cm]
\hline\\[-0.3cm]
\multicolumn{6}{l}{\parbox{7.7cm}{$^{\rm a}\!$\,{\scriptsize From\,unweighted\,observations\,for\,orbit\,C,\,weighted\,for\,{\it Best~Fit\/}.}}}\\[-0.45cm]
\end{tabular}}
\end{center}
\end{table}

Comparison of orbit C, the more realistic of the two sets in terms of the
comet's orbital period, with our optimum gravitational fit to the five
observations is presented in Table 5.  Marsden's (1989) residuals from
orbit C in this table have a tendency to become more positive toward both
ends of the observed arc in right ascension and to get more negative with
time in declination.  We tried to emulate the residuals from orbit C,
employing the set of truncated elements in Marsden's (1989) Table~V,
but succeeded to achieve this within about 4$^{\prime\prime}$ only in
declination.  Although we allowed for the rounding off of the perihelion
time, the residuals in right ascension came out systematically more
negative by about 12$^{\prime\prime}$ and all except the last one were
negative.  This effect is due apparently to the elements' truncation.

The residuals from our {\it Best Fit\/} solution, although clearly better than
those from orbit C, are not quite satisfactory either, suggesting that the
first observation may be inferior.  That should not be surprising, considering
that the comet was rather far from the plate center on this discovery exposure
(Section 3).  Unfortunately, the line of apsides deviates from the proper
direction by 0$^\circ\!$.571, which is unacceptably large.  Although the
individual angular elements of the {\it Best Fit\/} solution are burdened with
errors of up to several tenths of a degree, the correlation among them draws
the uncertainty in the orientation of the line of apsides down to only
hundredths of a degree.

Next, we compare Marsden's (1989) orbit D with the hybrid's orbit from Table 2.
 As with orbit C, we first tested whether the low-precision orbit D from
Marsden's Table~V is able to reproduce the residuals in his Table~VI.  Since
this exercise did not meet with success, we decided to reconstruct the
high-precision version of orbit D, not published by Marsden, by subtracting the
residuals in his Table~VI from the observed astrometric positions, retaining
his constraint on the reciprocal semimajor axis (0.01016~AU$^{-1}$).  This
approach was successful; the resulting orbit D is shown in Table 6 to
be exceedingly similar to the hybrid's orbit, the two sets of elements
differing mostly in the fourth or higher significant digit.  In
particular, we note that the lines of apsides agree to better than
0$^\circ\!\!$.01.  This correspondence is not fortuitous, as both sets
of elements represent perturbed versions of the orbits of two comets
observed 83~years apart, whose motions in space were virtually identical,
as already pointed out by Marsden (1967):\ C/1882~R1, used by us to
derive the hybrid's orbit; and C/1965~S1, employed by Marsden (1989) to
derive orbit D.

Table 6 also lists three sets of residuals: those left by orbit D from the
unweighted observations, in columns 2 and 3 [taken from Marsden's (1989)
Table~VI and marked M]; those left by orbit D from the weighted observations,
as derived by us\footnote{This fit required that the perihelion time be
increased by 0.0002 day relative to the tabulated value; otherwise the
residuals would be slightly greater.} (marked SK), in columns 4 and 5;
and those left by the hybrid's orbit from the weighted observations, in
columns 6 and 7.

\begin{table}[t]
\begin{center}
\vspace{0.1cm}
{\footnotesize {\bf Table 6}\\[0.08cm]
Marsden's (1989) Orbit D and Hybrid's Orbit (Equinox J2000).\\[0.07cm]
\begin{tabular}{l@{\hspace{0.4cm}}c@{\hspace{0.35cm}}c}
\hline\hline\\[-0.22cm]
Orbital element & Orbit D & Hybrid's orbit \\[0.06cm]
\hline\\[-0.2cm]
Perihelion time $t_\pi$ (1945 TT) & Dec.\,27.9798\rlap{$^{\rm a}$}
 & Dec.\,27.9801 \\
Argument of perihelion $\omega$ & $\;\:$67$^\circ\!$.8356
 & $\;\:$67$^\circ\!$.8514 \\
Longitude of ascending node $\Omega$ & 345$^\circ\!$.3571
 & 345$^\circ\!$.3753 \\
Orbit inclination $i$ & 141$^\circ\!$.5553 & 141$^\circ\!$.5607 \\
\hspace{3.05cm}(AU) & 0.007129 & 0.007123 \\[-0.33cm]
Perihelion distance $q \left\{ \raisebox{0ex}[0.3cm][0.3cm]{}
 \right.$\\[-0.39cm]
\hspace{3.05cm}(\Rssun) & 1.5315 & 1.5302 \\
Orbit eccentricity & 0.99992756 & 0.99992632 \\
Semimajor axis (AU) & 98.42 & 96.67 \\
Orbital period (yr) & 976.4 & 950.5 \\
Longitude of perihelion$^{\rm b}$, $L_\pi$ & $\;$282$^\circ\!$.84
 & $\;$282$^\circ\!$.84 \\
Latitude of perihelion$^{\rm b}$, $B_\pi$ & +35$^\circ\!$.16
 & +35$^\circ\!$.16 \\
Osculation epoch 1945 (TT) & Dec.\ 11.0 & Dec.\ 11.0 \\[0.06cm]
Mean residual$^{\rm c}$ \hspace{1.6cm} \rlap{{\hspace{0.4cm}}$\left\{
 \raisebox{0cm}[0.25cm][0.25cm]{} \right.$}
 & \raisebox{1.5ex}{$\pm$7$^{\prime\prime}\!\!$.63\,(M)}
 \hspace{-1.51cm}\raisebox{-1.5ex}{$\pm$8$^{\prime\prime}\!\!$.17\,(SK)}
 & $\pm$13$^{\prime\prime}\!\!$.75 \\[0.25cm]
\hline\\[-0.24cm]
\end{tabular}
\begin{tabular}{l@{\hspace{0.06cm}}r@{\hspace{0.4cm}}c@{\hspace{0.1cm}}c@{\hspace{0.38cm}}c@{\hspace{0.2cm}}c@{\hspace{0.43cm}}c@{\hspace{0.05cm}}c}
\multicolumn{8}{@{\hspace{0.1cm}}c}{Distribution of Residuals$^{\rm c}$,
 $O\!-\!C$} \\[0.1cm]
\hline \\[-0.2cm]
\multicolumn{2}{@{\hspace{-0.05cm}}c}{Time of}
& \multicolumn{2}{@{\hspace{-0.2cm}}c}{Orbit D\,(M)}
& \multicolumn{2}{@{\hspace{-0.2cm}}c}{Orbit D\,(SK)}
& \multicolumn{2}{@{\hspace{0cm}}c}{Hybrid's orbit} \\[-0.06cm]
\multicolumn{2}{@{\hspace{-0.05cm}}c}{observation}
& \multicolumn{2}{@{\hspace{-0.2cm}}c}{\rule[0.6ex]{1.8cm}{0.4pt}}
& \multicolumn{2}{@{\hspace{-0.2cm}}c}{\rule[0.6ex]{1.8cm}{0.4pt}}
& \multicolumn{2}{@{\hspace{0cm}}c}{\rule[0.6ex]{1.85cm}{0.4pt}} \\[-0.06cm]
\multicolumn{2}{@{\hspace{-0.05cm}}c}{1945 (UT)} & R.A. & $\;$Dec. & R.A.
  & $\;$Dec. & R.A. & $\;$Dec.\\[0.06cm]
\hline\\[-0.23cm]
Dec. & 11.04687 & $\!$+10$^{\prime\prime}\!\!$.8
     & $\,\;$+5$^{\prime\prime}\!\!$.1 & +12$^{\prime\prime}\!\!$.3
     & +6$^{\prime\prime}\!\!$.6 & $\:\!\!$+26$^{\prime\prime}\!\!$.4
     & $\;\;\,-$6$^{\prime\prime}\!\!$.3 \\ 
     & 12.04691 & $-$2.5 & \,$-$2.8 & $\;-$1.1 & $\!-$1.1 & $\!\!$+13.7
     & $-$12.1 \\
     & 13.07189 & $\!\!\!-$15.2 & $\!\!-$11.4 & $\!-$14.0 & $\!-$9.6
     & \,+1.2 & $-$18.7 \\
     & 14.07000 & +4.3 & $\,$+5.6 & $\;$+5.3 & $\!$+7.6 & $\!\!$+20.4
     & $\;\;\;\;\,$0.0 \\
     & 15.06885 & +2.8 & \,+2.5 & $\;$+3.6 & $\!$+4.5 & $\!\!$+18.4
     & $\;\:-$1.8 \\[0.04cm]
\hline\\[-0.28cm]
\multicolumn{8}{l}{\parbox{8.2cm}{$^{\rm a\!}$\,{\scriptsize To fit weighted
observations, \mbox{$t_\pi = {\rm Dec}.\,27.9800$}.}}}\\[-0.11cm]
\multicolumn{8}{l}{\parbox{8.2cm}{$^{\rm b\!}$\,{\scriptsize These
 ecliptical oordinates define the reference line of apsides.}}}\\[0cm]
\multicolumn{8}{l}{\parbox{8.2cm}{$^{\rm c\!}$\,{\scriptsize Entry (M) and
 {\vspace{-0.07cm}}columns 2--3 from unweighted observations; entry (SK) and
 columns 4--5 from weighted observations; {\vspace{-0.05cm}}residuals from
 hybrid's orbit are from weighted observations.}}}\\[0.2cm]
\end{tabular}}
\end{center}
\end{table}

Even though the residuals in Table 6 are fairly large, the ones left by
orbit D are much better than the strongly systematic ones left by the hybrid's
orbit.  From this comparison as well as from the difference in the mean
residual we infer that the orbital evolution of comet C/1945~X1, as a fragment
of its common parent with C/1882 R1 and C/1965 S1, was apparently more similar
to the orbital evolution of the latter than the former.  This is an important
though tentative conclusion, consistent with the fragmentation hierarchy of
the Kreutz system proposed by Sekanina \& Chodas (2004).  According to their
evolutionary model, C/1882 R1, C/1965 S1, and the precursor to C/1945 X1
separated from their common parent, possibly X/1106 C1, at the same time, some
18 days after perihelion.  Relative to C/1882 R1, the precursor moved in the
same direction as, and with a velocity only about 20\% lower than, C/1965~S1
(see also \mbox{Sekanina} \& Chodas 2002).  The subsequent separation of
C/1945~X1 from its precursor around 1700~{\small AD} notwithstanding, the
comet's orbit in 1945 should indeed resemble the orbit of C/1965~S1 to a
greater degree than that of C/1882~R1.

Since orbit D and the hybrid's orbit are so very similar, yet the residuals
they leave substantially differ, we tested whether this effect is due to
the minor differences in the angular elements or the orbital dimensions.
We replaced the semimajor axis 98.4 AU with 96.7 AU and noted that the
residuals from orbit~D in Table 6 did not change; the new orbit, D$^\prime$,
is now:\ \mbox{$t_\pi = {\rm 1945 \; December}\;27.9803$} TT,
\mbox{$q = 0.007126 \;{\rm AU} = 1.5309$ {\Rsun}}, \mbox{$e = 0.99992630$},
and for the equinox of J2000, \mbox{$\omega = 67^\circ\!.8377$},
\mbox{$\Omega = 345^\circ\!.3558$},~and \mbox{$i = 141^\circ\!.5570$}.~The
residuals in Table 6 reflect thus entire\-ly the differences of
up to 0$^\circ\!\!$.02 in the angular elements.

Before turning to nongravitational solutions, we take notice of a possibility
that one of the five astrometric observations is inferior and should be
discarded before computing any orbit.  Accordingly, we are now going to
search for solutions that could fit four observations.  Such solutions will
be referred to as ${\cal G}\{\Im\}$, where ${\cal G}$ stands for gravitational
and $\{\Im\}$ is a progression of reference numbers of the observations that
such solutions are based on; these numbers are listed in column 1 of Table 4.
The {\it Best Fit\/} solution presented in Table 5 can now be referred to
as ${\cal G}\{1,2,3,4,5\}$.  As already mentioned above, the discovery
position may be inferior, so that one of the tested four-observation
solutions is ${\cal G}\{2,3,4,5\}$.  The residuals from orbit D in Table 6
suggest that the third observation may be even worse than the first, raising
interest in the solution ${\cal G}\{1,2,4,5\}$.  Finally, the derivation of
the cataloged parabolic orbit implied that the fourth observation failed to
fit that solution (Marsden 1967), hence a need to test ${\cal G}\{1,2,3,5\}$.
The results of all three of these solutions are presented in Table 7.  No
orbits are being computed based on three observations only, as these are
regarded for our purposes as meaningless.
\begin{table*}[t]
\begin{center}
\vspace{0.1cm}
{\footnotesize {\bf Table 7}\\[0.08cm]
Comparison of Three Gravitational Solutions Based on\\Four Observations of
C/1945 X1 (Equinox J2000).\\[0.1cm]
\begin{tabular}{l@{\hspace{0.3cm}}c@{\hspace{0.3cm}}c@{\hspace{0.3cm}}c}
\hline\hline\\[-0.2cm]
Orbital element & ${\cal G}\{2,3,4,5\}$ & ${\cal G}\{1,2,4,5\}$
  & ${\cal G}\{1,2,3,5\}$ \\[0.1cm]
\hline\\[-0.22cm]
Perihelion time $t_\pi$ (1945 TT) & Dec.\,27.986 & Dec.\,27.979 & Dec.\ 27.983\\
Argument of perihelion $\omega$ & $\;\:$65$^\circ\!$.98
 & $\;\:$67$^\circ\!$.43 & $\:\;$74$^\circ\!$.00\\
Longitude of ascending node $\Omega$ & 342$^\circ\!$.84
 & 344$^\circ\!$.82 & 353$^\circ\!$.52 \\
Orbit inclination $i$ & 141$^\circ\!$.37 & 141$^\circ\!$.50 & 142$^\circ\!$.07\\
\hspace{3.05cm}(AU) & 0.007004 & 0.007120 & 0.007415 \\[-0.33cm]
Perihelion distance $q \left\{ \raisebox{0ex}[0.3cm][0.3cm]{}
 \right.$\\[-0.39cm]
\hspace{3.05cm}(\Rssun) & 1.505 & 1.530 & 1.593 \\
Orbit eccentricity & 0.99992755 & 0.99992635 & 0.99992329 \\
Longitude of perihelion $L_\pi$ & $\;$282$^\circ\!$.54
 & $\;$282$^\circ\!$.80 & $\;283^\circ\!$.49 \\
Latitude of perihelion $B_\pi$ & +34$^\circ\!$.77 & +35$^\circ\!$.09
 & +36$^\circ\!$.22 \\
Reference apsidal line's offset\rlap{$^{\rm a}$} & 0$^\circ\!$.462
 & 0$^\circ\!$.078 & 1$^\circ\!$.184\\
Osculation epoch 1945 (TT) & Dec.\ 11.0 &  Dec.\ 11.0 & Dec.\ 11.0 \\
Mean residual (weighted) & $\pm$5$^{\prime\prime}\!\!$.11
 & $\pm$1$^{\prime\prime}\!\!$.79
 & $\pm$1$^{\prime\prime}\!\!$.06 \\[0.05cm]
\hline\\[-0.24cm]
\end{tabular}
\begin{tabular}{l@{\hspace{0.06cm}}r@{\hspace{0.75cm}}c@{\hspace{0.3cm}}c@{\hspace{0.6cm}}c@{\hspace{0.3cm}}c@{\hspace{0.65cm}}c@{\hspace{0.3cm}}c}
\multicolumn{8}{@{\hspace{0.1cm}}c}{Distribution of Residuals$^{\rm b}$,
 $O \!-C\! $} \\[0.1cm]
\hline \\[-0.23cm]
\multicolumn{2}{@{\hspace{-0.3cm}}c}{Time of}
& \multicolumn{2}{@{\hspace{-0.4cm}}c}{${\cal G}\{2,3,4,5\}$}
& \multicolumn{2}{@{\hspace{-0.45cm}}c}{${\cal G}\{1,2,4,5\}$}
& \multicolumn{2}{@{\hspace{0cm}}c}{${\cal G}\{1,2,3,5\}$} \\[-0.05cm]
\multicolumn{2}{@{\hspace{-0.3cm}}c}{observation}
& \multicolumn{2}{@{\hspace{-0.4cm}}c}{\rule[0.6ex]{1.9cm}{0.4pt}}
& \multicolumn{2}{@{\hspace{-0.45cm}}c}{\rule[0.6ex]{1.9cm}{0.4pt}}
& \multicolumn{2}{@{\hspace{0cm}}c}{\rule[0.6ex]{1.9cm}{0.4pt}} \\
\multicolumn{2}{@{\hspace{-0.3cm}}c}{1945 (UT)} & R.A. & $\;$Dec. & R.A.
  & $\;$Dec. & R.A. & $\;$Dec.\\[0.07cm]
\hline\\[-0.2cm]
Dec. & 11.04687 &  \ldots\ldots & \ldots\ldots
     & +5$^{\prime\prime}\!\!$.3 & +1$^{\prime\prime}\!\!$.1
     & $\:\!\!-$3$^{\prime\prime}\!\!$.3 & $\;$+3$^{\prime\prime}\!\!$.3\\ 
     & 12.04691 & $\!\!\!$+25$^{\prime\prime}\!\!$.2
                & $\!\!\!$+12$^{\prime\prime}\!\!$.0
     & $\!-$8.2 & $\!-$7.4 & $\!\!$+4.9 & +2.7 \\
     & 13.07189 & $\!\!-$5.1 & $\!\!-$4.3 & \ldots\ldots & \ldots\ldots
     & $\!\!$+0.1 & $-$0.8 \\
     & 14.07000 & $\!\!$+4.3 & $\!\!$+5.5 & $\!$+0.1 & $\!$+1.3 & \ldots\ldots
     & \ldots\ldots \\
     & 15.06885 & $\!\!-$1.1 & $\!\!-$2.3 & $\!$+0.1 & $\!-$0.8 & $\!\!-$0.2
     & +0.3 \\[0.05cm]
\hline\\[-0.25cm]
\multicolumn{8}{l}{\parbox{8.5cm}{$^{\rm a\!}$\,{\scriptsize Reference line of
 apsides is defined by the hybrid's $L_\pi$ and $B_\pi$.}}}\\[-0.08cm]
\multicolumn{8}{l}{\parbox{8.5cm}{$^{\rm b\!}$\,{\scriptsize From weighted
observations.}}}\\[0cm]
\end{tabular}}
\end{center}
\end{table*}

Since a semimajor axis $a$ between~96.7~AU and 98.4 AU was shown to make no
difference, the four-observation solutions below are subjected to the same
constraint as the {\it Best Fit\/} solution in Table 5, namely, \mbox{$a
= 96.7$ AU} at an osculation epoch of 1945 December 26.0 TT.  The solutions
listed in Table 7 differ from one another considerably.  The orbits ${\cal
G}\{2,3,4,5\}$ and ${\cal G}\{1,2,3,5\}$ are both unacceptable on account
of their large offsets from the reference apsidal direction (given by the
hybrid's orbit and Marsden's orbit D).  The set ${\cal G}\{2,3,4,5\}$ is
in addition handicapped by the very large residuals left by the December
12 observation.  Even cursory inspection shows that the orbit
${\cal G}\{1,2,4,5\}$ is by far the most promising of the three, implying
that it is the December 13 observation that is inferior.  The offset from
the reference line of apsides, slightly less than 0$^\circ\!$.1, is reasonably
low though not completely satisfactory.  It is somewhat surprising that the
apparently inferior astrometric position is one of those taken with the
Metcalf refractor.  We would expect that the discarded data point should
be one of the first two observations, made with the Cooke camera.  Under the
circumstances, the solution ${\cal G}\{1,2,4,5\}$~can~at best be regarded
as marginally acceptable to approximate the orbital motion of C/1945 X1.

In any case, it is worth examining whether any meaningful refinement of
the orbit, an apsidal line in particular, can be achieved by incorporating
a nongravitational acceleration, which, if comparable to those affecting
the motions of the dwarf sungrazers, might be detectable over a span of
four days.  Besides, such alternative solutions should prove beneficial
to estimating a range of uncertainties in the comet's orbital motion
(Section 7).

\subsection{Standard Nongravitational Solutions}
In order to find out whether there is evidence for nongravitational
effects in the motion of comet C/1945~X1 in the meager set of astrometric
data available, we begin with a standard formalism of Marsden et al.\
(1973), based on a water-ice sublimation model.  The erosion-driven
nongravitational accelerations in the three cardinal directions, i.e.,
in the radial (away~from the Sun), {\bf R}, transverse, {\bf T}, and
normal, {\bf N}, directions~of~a right-handed {\bf RTN} coordinate
system that is referred~to the comet's orbital plane, are in this
formalism expressed by
\begin{equation}
\left[ \!\!
\begin{array}{c}
a_{\rm R}(r) \\
a_{\rm T}(r) \\
a_{\rm N}(r)
\end{array}
\!\! \right] = \left[ \!\!
\begin{array}{c}
A_1 \\
A_2 \\
A_3
\end{array}
\!\! \right] \! \cdot g_{\rm std}(r),
\end{equation}
where $g_{\rm std}(r)$ is the standard law employed by Marsden et al.\ (1973),
approximating the sublimation rate of water ice from an isothermal spherical
nucleus and normalized so that \mbox{$g_{\rm std}(1\;{\rm AU}) = 1$},
\begin{equation}
g_{\rm std}(r) = \alpha \left( \! \frac{r}{r_0} \! \right)^{\!\!-m} \!\!
\left[ 1 \!+\! \left( \! \frac{r}{r_0} \! \right)^{\!\!n} \right]^{-k}\!\!\!.
\end{equation}
Here \mbox{$m \!=\! 2.15$}, \mbox{$n \!=\! 5.093$}, \mbox{$nk \!=\! 23.5$}, a
scaling distance \mbox{$r_0 \!=\! 2.808$ AU}, and a normalization constant
\mbox{$\alpha \!=\! 0.1113$}.  The parameters $A_1$, $A_2$, $A_3$ are,
respectively, the radial, transverse, and normal components of the
nongravitational acceleration at 1 AU from the Sun; their units usually
employed in orbital studies are 10$^{-8}\!$ AU day$^{-2}$.  Since the vector
of the acceleration due to the generally sunward-directed sublimation points
away from the~Sun, physically meaningful values of $A_1$ should always be
positive.  The law (9) has over the past 40~years served admirably in countless
orbit-determination applications to comets with perihelia typically several
tenths of AU.

The numerical procedure that we apply next follows the method described in
Paper~1.  Briefly, since the direction of the line of apsides was found to be
a function~of the nongravitational acceleration, a minimum offset from the
reference, or nominal, apsidal line (defined here by $L_\pi$ and $B_\pi$ for
the hybrid's orbit or orbit D; see Table 6) serves as the constraint that
determines the nongravitational acceleration's most probable magnitude.

Similarly to our notation for{\vspace{-0.1cm}} the gravitational runs, we
refer to these nongravitational solutions by ${\cal N}_{\rm std}^{\,\bf
(\!X\!)}\!\{\Im\}$, where $\{\Im\}$ is again a progression{\vspace{-0.09cm}}
of reference numbers~of the employed observations from Table 4, while ${\cal
N}_{\rm std}^{\,\bf (\!X\!)}$ denotes, on the one hand, Marsden et al.'s
(1973) formalism of accounting for the nongravitational effect, with the
standard law (9) describing the variations with heliocentric distance; and,
on the other hand, one of the two versions applied:\ either solving for the
radial component of the acceleration with the parameter $A_1$ when \mbox{$\bf
X = \bf R$};~or for the normal component with $A_3$ when \mbox{$\bf X =
\bf N$}.  Whenever we attempted to solve for both components, the run aborted.
We recall that the radial component (\mbox{$A_1 \!>\! 0$}) dominates the
magnitude of the nongravitational acceleration in the motions of the cataloged
comets in nearly-parabolic orbits (Marsden \& Williams 2008), while the normal
component ($A_3$) was found to contribute significantly to the magnitude of the
acceleration that affects the motions of the dwarf Kreutz sungrazers (Paper 1).

Table 8 summarizes the most important results from the runs based on Marsden
et al.'s (1973) standard~formalism.  Altogether we computed eight such
solutions~for four different observational sets~$\{\Im\}$,{\vspace{-0.07cm}}
the same ones~as~before:\ \mbox{${\cal N}_{\rm std}^{\,\bf (\!R\!)}\{1,2,3,4,5\}$},
\mbox{${\cal N}_{\rm std}^{\,\bf (\!R\!)}\{2,3,4,5\}$},{\vspace{-0.07cm}}~\mbox{${\cal N}_{\rm std}^{\,\bf (\!R\!)}\{1,2,4,5\}$},
\mbox{${\cal N}_{\rm std}^{\,\bf (\!R\!)}\{1,2,3,5\}$}; and
similarly{\vspace{-0.11cm}} with $\bf N$ instead of $\bf R$.~Both the
$\bf R$ and the $\bf N$ versions{\vspace{-0.12cm}} of the~\mbox{${\cal N}_{\rm
std}^{\,\bf (\!X\!)}\{1,2,3,4,5\}$}~and ${\cal N}_{\rm
std}^{\,\bf (\!X\!)}\{1,2,3,5\}$ runs are inferior to their respective
gravitational solutions (Tables 5 and 7) in terms of the{\vspace{-0.015cm}}
mean residual:\ $\pm$6$^{\prime\prime}\!$.42 ($\bf R$) and $\pm$6$^{\prime
\prime} \!$.68 ($\bf N$) against~$\pm$6$^{\prime\prime}\!$.17~in the case of
the $\{1,2,3,4,5\}$ solution; and $\pm$3$^{\prime\prime}\!$.03 ($\bf R$) and
$\pm$2$^{\prime\prime}\!$.61 ($\bf N$) against $\pm$1$^{\prime\prime}\!$.06
in the case of the \{1,2,3,5\} solution.  In addition, for these two sets of
solutions the $\bf R$-version parameters $A_1$ come out to be negative and
therefore physically meaningless.

\begin{table}[t]
\begin{center}
\vspace{0.12cm}
{\footnotesize {\bf Table 8}\\[0.1cm]
Comparison of the Nongravitational Parameters\\and Test Results for Eight
Orbital Solutions\\Based on Marsden et al.'s formalism.\\[0.2cm]
\begin{tabular}{l@{\hspace{0.65cm}}c@{\hspace{0.6cm}}c@{\hspace{0.35cm}}c}
\hline\hline\\[-0.2cm]
                 &                     & Apsidal- & Mean \\[-0.05cm]
Nongravitational & Parameter$^{\rm a}$ & \,line$^{\rm b}$ & residual \\[-0.05cm]
solution         & $A_1$ or $A_3$ & offset & (weighted) \\[0.1cm]
\hline\\[-0.15cm]
${\cal N}_{\rm std}^{\,\bf (\!R\!)}\{1,2,3,4,5\}$ & $-$0.82\,$\pm$\,0.17
 & 0$^\circ\!\!$.116 & $\pm$6$^{\prime\prime}\!\!$.42 \\[0.05cm]
${\cal N}_{\rm std}^{\,\bf (\!R\!)}\{2,3,4,5\}$   & +0.67\,$\pm$\,0.18
 & 0.118 & $\pm$5.05 \\[0.05cm]
${\cal N}_{\rm std}^{\,\bf (\!R\!)}\{1,2,4,5\}$   & +0.113\,$\pm$\,0.019
 & 0.013 & $\pm$1.73 \\[0.05cm]
${\cal N}_{\rm std}^{\,\bf (\!R\!)}\{1,2,3,5\}$   & $-$1.70\,$\pm$\,0.40
 & 0.270 & $\pm$3.03 \\[0.05cm]
${\cal N}_{\rm std}^{\,\bf (\!N\!)}\{1,2,3,4,5\}$ & $-$1.87\,$\pm$\,0.23
 & 0.073 & $\pm$6.68 \\[0.05cm]
${\cal N}_{\rm std}^{\,\bf (\!N\!)}\{2,3,4,5\}$   & +1.54\,$\pm$\,0.30
 & 0.085 & $\pm$4.88 \\[0.05cm]
${\cal N}_{\rm std}^{\,\bf (\!N\!)}\{1,2,4,5\}$   & +0.234\,$\pm$\,0.018
 & 0.006 & $\pm$1.72 \\[0.05cm]
${\cal N}_{\rm std}^{\,\bf (\!N\!)}\{1,2,3,5\}$   & $-$4.18\,$\pm$\,0.65
 & 0.212 & $\pm$2.61 \\[0.1cm] 
\hline\\[-0.22cm]
\multicolumn{4}{l}{\parbox{7.8cm}{$^{\rm a}$\,{\scriptsize In units of
 10$^{-5}$AU\,day$^{-2}$.}}} \\[-0.01cm]
\multicolumn{4}{l}{\parbox{7.8cm}{$^{\rm b\!}$\,{\scriptsize Offset from
 direction of the reference line of apsides (Table 6).}}} \\[0.1cm]
\end{tabular}}
\end{center}
\end{table}

\begin{table}[hb]
\begin{center}
\vspace{0.3cm}
{\footnotesize {\bf Table 9}\\[0.12cm]
Orbital Solutions ${\cal N}_{\rm std}^{\,\bf (\!R\!)}\{1,2,4,5\}$ and
${\cal N}_{\rm std}^{\,\bf (\!N\!)}\{1,2,4,5\}$\\[0.05cm]
 (Equinox J2000).\\[0.12cm]
\begin{tabular}{l@{\hspace{0.05cm}}c@{\hspace{0.25cm}}c}
\hline\hline\\[-0.2cm]
Orbital element & ${\cal N}_{\rm std}^{\,\bf (\!R\!)}\!\{1,2,4,5\}$
 & ${\cal N}_{\rm std}^{\,\bf (\!N\!)}\!\{1,2,4,5\}$ \\[0.1cm]
\hline\\[-0.2cm]
Perihelion time $t_\pi$ (1945 TT) & Dec.\,27.97928 & Dec.\,27.97938 \\
Argument of perihelion $\omega$ & $\;\:$67$^\circ\!$.910
 & $\;\:$67$^\circ\!$.883 \\
Longitude of ascending node $\Omega$ & 345$^\circ\!$.437
 & 345$^\circ\!$.412 \\
Orbit inclination $i$ & 141$^\circ\!$.572 & 141$^\circ\!$.565 \\
\hspace{3.05cm}(AU) & 0.0071167 & 0.0071250 \\[-0.33cm]
Perihelion distance $q \left\{ \raisebox{0ex}[0.3cm][0.3cm]{}
 \right.$\\[-0.39cm]
\hspace{3.05cm}(\Rssun) & 1.5287 & 1.5307 \\
Orbit eccentricity & 0.99992638 & 0.99992629 \\
Longitude of perihelion $L_\pi$ & $\;$282$^\circ\!$.82
 & $\;$282$^\circ\!$.83 \\
Latitude of perihelion $B_\pi$ & +35$^\circ\!$.16 & +35$^\circ\!$.16 \\
Reference apsidal-line offset  & 0$^\circ\!$.0133 & 0$^\circ\!$.0060 \\
Parameter $A_1$\,(10$^{-5}$AU\,day$^{-2})$ & +0.113$^{\rm a}$
 & \,\ldots\ldots \\
Parameter $A_3$\,(10$^{-5}$AU\,day$^{-2})$ & \,\ldots\ldots
 & +0.234$^{\rm b}$ \\
Osculation epoch 1945 (TT) & Dec.\ 11.0  &  Dec.\ 11.0 \\
Mean residual (weighted) & $\pm$1$^{\prime\prime}\!\!$.73
                         & $\pm$1$^{\prime\prime}\!\!$.72 \\[0.05cm]
\hline\\[-0.24cm]
\end{tabular}
\begin{tabular}{l@{\hspace{0.07cm}}r@{\hspace{1.1cm}}c@{\hspace{0.6cm}}c@{\hspace{1.1cm}}c@{\hspace{0.25cm}}c}
\multicolumn{6}{@{\hspace{0.1cm}}c}{Distribution of Residuals $O\!-\!C$} \\[0.1cm]
\hline \\[-0.24cm]
\multicolumn{2}{@{\hspace{-0.75cm}}c}{Time of}
& \multicolumn{2}{@{\hspace{-0.85cm}}c}{${\cal N}_{\rm std}^{\,\bf
 (\!R\!)}\!\{1,2,4,5\}$}
& \multicolumn{2}{@{\hspace{-0.35cm}}c}{${\cal N}_{\rm std}^{\,\bf
 (\!N\!)}\!\{1,2,4,5\}$} \\[-0.05cm]
\multicolumn{2}{@{\hspace{-0.75cm}}c}{observation}
& \multicolumn{2}{@{\hspace{-0.85cm}}c}{\rule[0.6ex]{2.15cm}{0.4pt}}
& \multicolumn{2}{@{\hspace{-0.35cm}}c}{\rule[0.6ex]{2.15cm}{0.4pt}} \\
\multicolumn{2}{@{\hspace{-0.75cm}}c}{1945 (UT)} & R.A. & Dec. & R.A.
  & Dec. \\[0.05cm]
\hline\\[-0.24cm]
Dec. & 11.04687 & +4$^{\prime\prime}\!\!$.9 & $\!$+2$^{\prime\prime}\!\!$.0
 & +4$^{\prime\prime}\!\!$.9 & +1$^{\prime\prime}\!\!$.6 \\ 
     & 12.04691 & $-$8.2 & $-$7.3 & $-$8.0 & $-$7.2 \\
     & 14.07000 & +0.3 & +1.0 & +0.2 & +1.1 \\
     & 15.06885 & $-$0.1 & $-$0.6 & $\;\;\,$0.0 & $-$0.7 \\[0.05cm]
\hline\\[-0.28cm]
\multicolumn{6}{l}{\parbox{8.25cm}{$^{\rm a}${\scriptsize With a mean error
 of $\pm 0.019 \times 10^{-5}$AU\,day$^{-2}$ (Table 8).}}}\\[-0.08cm]
\multicolumn{6}{l}{\parbox{8.25cm}{$^{\rm b}${\scriptsize With a mean error
 of $\pm 0.018 \times 10^{-5}$AU\,day$^{-2}$ (Table 8).}}}\\[-0.42cm]
\end{tabular}}
\end{center}
\end{table}

Even though the set $\{2,3,4,5\}$ avoids the misfortunes of the sets
$\{1,2,3,4,5\}$ and $\{1,2,3,5\}$, its offsets from the reference line of
apsides fail to improve over the corresponding offsets from $\{1,2,3,4,5\}$;
in addition, the residuals left by the December 12 observation are comparable
to those from the equivalent gravitational solution (Table~7) and entirely
unacceptable.  In summary, the standard nongravitational solutions based on
the set $\{1,2,4,5\}$ are by far the best, in terms of both the offset from
the reference apsidal line and the mean residual.  Also, both $A_1$ and
$A_3$ of the $\{1,2,4,5\}$ solutions come out in Table~8 to be almost or
just about one order of magnitude smaller than those of the poor solutions.

As the final comments we note that (i) relatively to the gravitational fit
in Table 7, both nongravitational solutions based on the set $\{1,2,4,5\}$
improve the mean residual only marginally (from $\pm$1$^{\prime\prime}\!$.79
to $\pm$1$^{\prime\prime}\!$.72/1$^{\prime\prime}\!$.73), but reduce the
offset from the reference apsidal line considerably (from 0$^\circ\!$.078 to
0$^\circ\!$.013/0$^\circ\!$.006); (ii) both $A_1$ and $A_3$ appear to be
fairly well determined (with the relative errors of $\pm$8--17\%) and on the
{\vspace{-0.03cm}}same order of magnitude, $\sim$10$^{-6}$\,AU day$^{-2}$, as
the parameters of the normal component of the nongravitational acceleration
of the dwarf sungrazers C/2009 L5 and C/2006 J9 in Table 4 of Paper 1; and
(iii)~these parameters are one order of magnitude greater than the peak values
of $A_1$ for the cataloged comets in nearly-parabolic orbits and at least two
orders of magnitude greater than the typical values of $A_1$ for such comets
(Marsden \& Williams 2008).  In summary, there is some --- though, due to
the limited~data, not overwhelming --- evidence that in terms of the
non\-gravitational effects in its orbital motion, C/1945~X1 may share the
properties of some dwarf Kreutz sungrazers.

\subsection{Modified Nongravitational Solutions}
The EXORB7 orbit-determination code employed by the second author allows
one to vary arbitrarily all five parameters of the nongravitational law (9)
--- the exponents $m$, $n$, $k$, the scaling distance $r_0$, and the
normalization constant $\alpha$.  This option facilitates a more robust
orbital experimentation, when the standard nongravitational model of
Marsden et al.\ (1973) fails to provide satisfactory results.  In Paper 1 we
gained some experience with what we hereafter refer to as a {\it modified\/}
nongravitational law $g_{\rm mod}(r;r_0)$, which is given by Equation (9),
but while the three exponents remain unchanged from the standard law, the
scaling distance varies broadly, always entailing a change in $\alpha$
as well.

The physics behind this sublimation-law experimentation involves a parallelism
between the scaling distance $r_0$ and the {\it snow line\/}, a boundary of
the zone in vacuum (or near vacuum), beyond which it is cold enough for a
volatile substance to exist only in the solid phase.  Calibrated by water
ice, for which in the isothermal model \mbox{$r_0 \simeq 2.8$ AU}, this
distance depends in the first approximation on the effective latent heat
of sublimation, $L$ (in cal mol$^{-1}$), of the sublimating species:
\begin{equation}
r_0 \, \dot{=} \left( \frac{\rm const}{L} \right)^{\!2} \!,
\end{equation}
where $r_0$ is expressed in AU and the constant{\vspace{-0.03cm}} is equal
to 19\,100~AU$^{\frac{1}{2}}$\,cal\,mol$^{-1}$.  Marsden et al.\ (1973)
showed that in a plot of log\,(normalized sublimation rate) against
log\,(heliocentric distance) the value of $r_0$ shifts the curve left or
right along the axis of heliocentric distance $r$. Hence, by properly
choosing $r_0$, the erosion-driven nongravitational effects in
the orbital motion of any comet can approximately be expressed by a
universal curve of log\,(normalized sublimation rate) against $\log(r/r_0)$.

Highly refractory materials, such as metals or silicates, can sublimate only
very close to the Sun, so that their snow line and scaling distance are much
smaller than 1~AU.  For example, for forsterite, the Mg end-member of the
olivine solid solution system, the latent heat of sublimation is
130\,000~cal mol$^{-1}$ (Nagahara et al.\ 1994), so that its \mbox{$r_0
\simeq 0.02$ AU}.

In limiting cases the empirical equation (9) offers the following expressions
for variations in the sublimation rates (equally applying to the modified and
standard laws):
\begin{eqnarray}
\lim_{x \rightarrow 0} g_{\rm mod}(r; r_0) & \sim & r^{\!-2.15}\!,
 \nonumber \\[0cm]
\lim_{x \rightarrow \infty} g_{\rm mod}(r; r_0) & \sim &
 r^{\!-25.65}\!.
\end{eqnarray}
where \mbox{$x = r/r_0$}.  The first limit approaches an extreme scenario
in which all incident solar energy is spent on sublimation, while the second
limit crudely approximates the other extreme, when the energy is spent
entirely on heating the object{\vspace{-0.03cm}} (and increasing its
thermal reradiation).  The actual limiting expressions are $r^{-2}$ for
\mbox{$r/r_0 \rightarrow 0$} and
\begin{equation}
\gamma_{\rm mod}(r; r_0) \sim \exp \! \left( \! -\frac{L}{RT}\right) \;\;\;
 {\rm for} \; r/r_0 \rightarrow \infty,
\end{equation}
where the sublimation heat $L$ is related to $r_0$ by Equation (10), $R$
is the gas constant, and $T$ is the temperature at heliocentric distance
$r$.  The differences relative to (11) are due to the approximate nature of
the law (9), which is responsible for a peculiar feature when applied to
the orbital motion of C/1945 X1 (see below).  To illustrate it, we begin by
introducing a local slope $\zeta$ of a modified nongravitational law between
$r$ and $r + \Delta r$; similarly to (11), it is

\begin{equation}
\lim_{\Delta r \rightarrow 0} g_{\rm mod}(r; r_0) \sim r^{-\zeta},
\end{equation}
where
\begin{equation}
\zeta(r; r_0) = -\, \frac{\partial \ln g_{\rm mod}(r; r_0)}{\partial \ln(r/r_0)}
 = m + \frac{nk}{1 + (r/r_0)^{-n}}.
\end{equation}
The limits on $\zeta$ are those given in (11).  Since $r_0$ is nearly 3 AU in
the standard law, the contribution to the integrated nongravitational effect
from \mbox{$r \!\gg\! r_0$} is always negligible and the exact shape of the
standard nongravitational law at heliocentric distances \mbox{$r \!\gg\! r_0$}
is unimportant.  However, when $r_0$ in a modified law is considerably smaller
than that of the standard law, the power-law approximation to the exponential
law for the sublimation rate beyond $r_0$ could play a major role for comets
whose perihelion distances $q$ are much greater than $r_0$.

\begin{figure}[t]
\vspace{-2.2cm}
\hspace{1.55cm}
\centerline{
\scalebox{0.63}{
\includegraphics{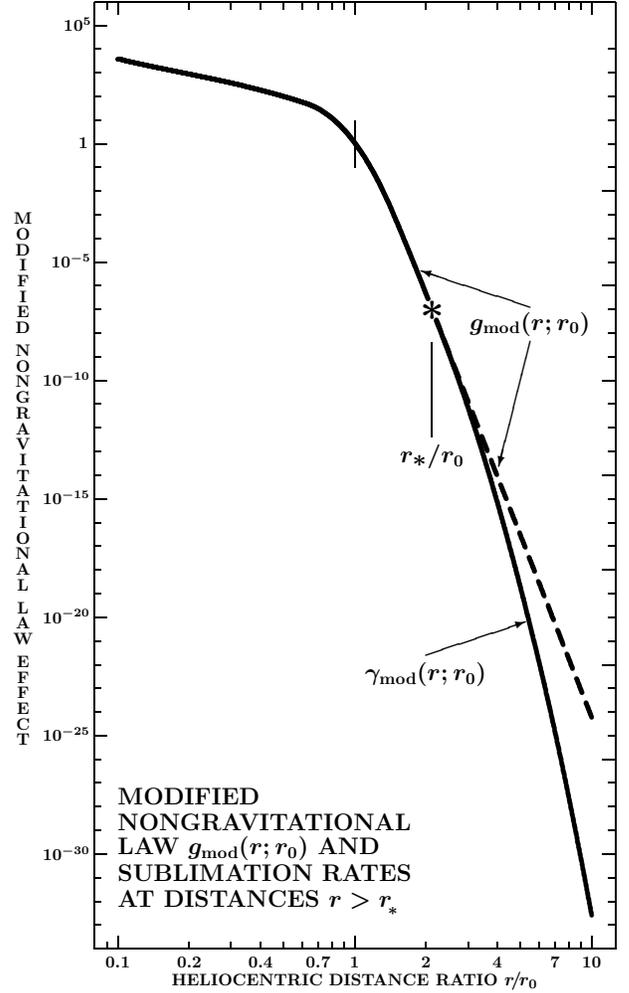}}} 
\vspace{-3.55cm}
\caption{The nongravitational effects in the orbital motion of a comet
derived from the modified law $g_{\rm mod}(r; r_0)$ applicable to an
arbitrary scaling distance{\vspace{-0.05cm}} $r_0$.  At heliocentric
distances \mbox{$r > r_{_{\mbox{\boldmath $\displaystyle \ast$}}}$} the
law $\gamma_{\rm mod}(r; r_0)$ matches the exponential variations in the
sublimation rate better than $g_{\rm mod}(r; r_0)$. The laws $g_{\rm
mod}(r; r_0)$ and $\gamma_{\rm mod}(r; r_0)$ have equal{\vspace{-0.05cm}}
logarithmic slopes at $r_{_{\mbox{\boldmath $\displaystyle \ast$}}}$.  The
slope $\zeta$ of $g_{\rm mod}(r; r_0)$ at \mbox{$r > r_{_{\mbox{\boldmath
$\displaystyle \ast$}}}$} is practically constant.{\vspace{0.3cm}}}
\end{figure}

To find out at what heliocentric distance $r_{_{\mbox{\boldmath $\ast$}}}$
{\vspace{-0.05cm}}does a modified law begin to deviate from the sublimation
law $\gamma_{\rm mod}(r; r_0)$, we first define the local slope $\xi$ of
$\gamma_{\rm mod}(r; r_0)$ similarly to (14):
\begin{equation}
\xi(r; r_0) = -\,\frac{\partial \ln \gamma_{\rm mod}(r; r_0)}{\partial
 \ln(r/r_0)} = \frac{L \sqrt{r}}{2RT_0} = \frac{L\sqrt{r_0}}{2RT_0}
 \sqrt{\frac{r}{r_0}},
\end{equation}
where \mbox{$T(r) = T_0/\sqrt{r}$} at the heliocentric distances of
extremely low sublimation rates; for the isothermal model, \mbox{$T_0
= 278.3$\,K}.  Since, from Equation (10), $L \sqrt{r_0}$ is a constant
and \mbox{$L \sqrt{r_0}/ RT_0 = 34.54$}, the point of deviation of the
modified law $g_{\rm mod}(r; r_0)$ from the exponential sublimation law,
given by the condition of equal slopes, \mbox{$\xi(r_{_{\mbox{\boldmath
$\ast$}}}; r_0) = \zeta(r_{_{\mbox{\boldmath $\ast$}}}; r_0)$}, determines
$r_{_{\mbox{\boldmath $\ast$}}}$  by comparing (14) with (15).  After
inserting the numerical values,
\begin{equation}
r_{_{\mbox{\boldmath $\ast$}}} = 2.12 \, r_0.
\end{equation}
Less than a millionth part of the Sun's incident energy at the distance
$r_{_{\mbox{\boldmath $\displaystyle \ast$}}}$ is spent on the sublimation.

Requiring that $g_{\rm mod}(r_{_{\mbox{\boldmath $\ast$}}};r_0) = \gamma_{\rm
mod}(r_{_{\mbox{\boldmath $\ast$}}};r_0)$ and writing the exponential
sublimation law in the form
\begin{equation}
\gamma_{\rm mod}(r; r_0) = \beta \, \exp \! \left[ 34.54 \left(1 -
 \sqrt{\frac{r}{r_0}} \,\right) \right],
\end{equation}
we have
\begin{equation}
\beta = 0.0267 \alpha.
\end{equation}
The relationship between $g_{\rm mod}(r; r_0)$ and $\gamma_{\rm mod}(r;
r_0)$ is displayed in Figure 4.  At present, the $\gamma _{\rm mod}(r;
r_0)$ law is not incorporated into the orbit-determination code that we
employ, as in most cases it would make hardly any difference
numerically.  However, the approximation of $\gamma_{\rm mod}(r; r_0)$
by $g_{\rm mod}(r; r_0)$ at \mbox{$r > r_{_{\mbox{\boldmath $\ast$}}}$}
(i)~prevents one from testing nongravitational laws with a variable
exponent $\zeta$ at these heliocentric distances and (ii)~yields
identical results from all solutions based on the scaling distances
$r_0$ that are by more than 2.12\,$r_0$ smaller than the least
heliocentric distance at which the comet is observed, as seen from Figure 4.

In Paper 1 we employed the modified-law paradigm to great advantage.  We
determined for all eight in-depth examined dwarf sungrazers that the
astrometric positions were fitted better (and the offsets from the reference
line of apsides came out to be smaller) when the nongravitational acceleration
affecting their orbital motions was assumed to vary with heliocentric distance
$r$ much more steeply than prescribed by the standard law $g_{\rm std}(r)$.
The orbital motions of six out of the eight cases were fitted best with
\mbox{$r_0 \!<\! 0.13$ AU}, which corresponds to \mbox{$L \mbox{\gapeq}
50\,000$ cal mol$^{-1}$}, and for none of the eight was $r_0$ higher than
$\sim$2.1~AU.  The heliocentric distances of the dwarf Kreutz sungrazers in
Paper 1 varied between 0.037 and 0.068 AU (or between 8 and 15 {\Rsun}), so
that these objects were considerably closer to the Sun than C/1945~X1 when
under observation from Boyden.

\begin{table}[t]
\begin{center}
 \vspace{0.13cm}
{\footnotesize {\bf Table 10}\\[0.08cm]
Orbital Solutions ${\cal N}_{\rm mod}^{\,\bf (\!R\!)}[0.05]\{\Im_0\}$
and ${\cal N}_{\rm mod}^{\,\bf (\!N\!)}[0.05]\{\Im_0\}$\\[0.07cm]
 (Equinox J2000).\\[0.05cm]
\begin{tabular}{l@{\hspace{-0.15cm}}c@{\hspace{0.1cm}}c}
\hline\hline\\[-0.24cm]
Orbital element & ${\cal N}_{\rm mod}^{\,\bf (\!R\!)}[0.05]\!\{\Im_0\}$
 & ${\cal N}_{\rm mod}^{\,\bf (\!N\!)}[0.05]\!\{\Im_0\}$ \\[0.1cm]
\hline\\[-0.2cm]
Perihelion time $t_\pi$ (1945 TT) & Dec.\,27.97937 & Dec.\,27.97940 \\
Argument of perihelion $\omega$ & $\;\:$67$^\circ\!$.876
 & $\;\:$67$^\circ\!$.869 \\
Longitude of ascending node $\Omega$ & 345$^\circ\!$.403
 & 345$^\circ\!$.397 \\
Orbit inclination $i$ & 141$^\circ\!$.562 & 141$^\circ\!$.561 \\
\hspace{3.05cm}(AU) & 0.0071256 & 0.0071279 \\[-0.33cm]
Perihelion distance $q \left\{ \raisebox{0ex}[0.3cm][0.3cm]{}
 \right.$\\[-0.39cm]
\hspace{3.05cm}(\Rssun) & 1.5308 & 1.5313 \\
Orbit eccentricity & 0.99992629 & 0.99992626 \\
Longitude of perihelion $L_\pi$ & $\;$282$^\circ\!$.83
 & $\;$282$^\circ\!$.84 \\
Latitude of perihelion $B_\pi$ & +35$^\circ\!$.16 & +35$^\circ\!$.16 \\
Reference apsidal-line offset  & 0$^\circ\!$.0055 & 0$^\circ\!$.0034 \\
Parameter $A_1$\,(10$^{-10}$AU\,day$^{-2})$ & +0.307$^{\rm a}$
 & \,\ldots\ldots \\
Parameter $A_3$\,(10$^{-10}$AU\,day$^{-2})$ & \,\ldots\ldots
 & +0.877$^{\rm b}$ \\
Osculation epoch 1945 (TT) & Dec.\ 11.0  &  Dec.\ 11.0 \\
Mean residual (weighted) & $\pm$1$^{\prime\prime}\!\!$.66
                         & $\pm$1$^{\prime\prime}\!\!$.66 \\[0.05cm]
\hline\\[-0.26cm]
\end{tabular}
\begin{tabular}{l@{\hspace{0.07cm}}r@{\hspace{1.1cm}}c@{\hspace{0.6cm}}c@{\hspace{1.1cm}}c@{\hspace{0.2cm}}c}
\multicolumn{6}{@{\hspace{0.1cm}}c}{Distribution of Residuals
 $O\!-\!C$} \\[0.08cm]
\hline \\[-0.24cm]
\multicolumn{2}{@{\hspace{-0.75cm}}c}{Time of}
& \multicolumn{2}{@{\hspace{-0.85cm}}c}{$\,{\cal N}_{\rm mod}^{\,\bf
 (\!R\!)}[0.05]\{\Im_0\}$}
& \multicolumn{2}{@{\hspace{-0.35cm}}c}{$\,{\cal N}_{\rm mod}^{\,\bf
 (\!N\!)}[0.05]\{\Im_0\}$} \\[-0.05cm]
\multicolumn{2}{@{\hspace{-0.75cm}}c}{observation}
& \multicolumn{2}{@{\hspace{-0.85cm}}c}{\rule[0.6ex]{2.15cm}{0.4pt}}
& \multicolumn{2}{@{\hspace{-0.35cm}}c}{\rule[0.6ex]{2.15cm}{0.4pt}}\\[-0.03cm]
\multicolumn{2}{@{\hspace{-0.75cm}}c}{1945 (UT)} & R.A. & Dec. & R.A.
  & Dec. \\[0.05cm]
\hline\\[-0.24cm]
Dec. & 11.04687 & +4$^{\prime\prime}\!\!$.7 & $\!$+2$^{\prime\prime}\!\!$.0
 & +4$^{\prime\prime}\!\!$.8 & +1$^{\prime\prime}\!\!$.7 \\ 
     & 12.04691 & $-$8.0 & $-$6.9 & $-$7.9 & $-$6.9 \\
     & 14.07000 & +0.3 & +0.9 & +0.3 & +1.0 \\
     & 15.06885 & $-$0.1 & $-$0.5 & $\;\;\,$0.0 & $-$0.6 \\[0.03cm]
\hline\\[-0.28cm]
\multicolumn{6}{l}{\parbox{8.25cm}{$^{\rm a}${\scriptsize With a mean error
 of $\pm 0.021 \times 10^{-10}$AU\,day$^{-2}$.}}}\\[-0.08cm]
\multicolumn{6}{l}{\parbox{8.25cm}{$^{\rm b}${\scriptsize With a mean error
 of $\pm 0.046 \times 10^{-10}$AU\,day$^{-2}$.}}}\\[0.21cm]
\end{tabular}}
\end{center}
\end{table}

For C/1945 X1, the modified-law nongravitational solutions with \mbox{$r_0
\ll 0.2$ AU} fit most satisfactorily the preferred set of four observations,
\mbox{$\{\Im_0\} = \{1, 2, 4, 5\}$}.  This match is better than that from
the standard-law solutions listed in Table 9.  We refer to these modified-law
solutions as ${\cal N}_{\rm mod}^{\,(\!\bf X\!)} [r_0]\{\Im_0\}$, where {\bf
X} has the same meaning as before and $r_0$ is in AU.  Because the comet was
not observed closer to the Sun than 0.6 AU, all these solutions are identical,
as pointed out more generally under (ii) below Equation (18).  We confirm this
result by running solutions for $r_0$ equal {\vspace{-0.1cm}}to 0.01, 0.05,
and 0.10 AU.  We~then select, as examples, the solutions{\vspace{-0.14cm}}
${\cal N}_{\rm mod}^{\,(\!\bf R\!)}[0.05]\{\Im_0\}$~and ${\cal N}_{\rm
mod}^{\,(\!\bf N\!)}[0.05]\{\Im_0\}$ and present them in Table 10; relative
to the results based on the standard law, the offsets from the reference
line of apsides now dropped, respectively, from 0$^\circ\!$.013 to
0$^\circ\!$.005 and from 0$^\circ\!$.006 to 0$^\circ\!$.003, while the
weighted mean residuals dropped, respectively, from
$\pm$1$^{\prime\prime}\!$.73 and $\pm$1$^{\prime\prime}\!$.72 down to
$\pm$1$^{\prime\prime}\!$.66 in both {\bf R} and {\bf N}.

The nongravitational parameters $A_1$ and $A_3$ are now determined with
relative errors of only $\pm$5--7\%.  Their apparent discrepancy of nearly
five orders of magnitude compared to those of the standard law is due
entirely to the very different steepness of the two laws' slopes.  Since
the comet was observed between 0.718 AU from the Sun (on December 11) and
0.598 AU from the Sun (on December 15), the magnitudes of the effective
nongravitational accelerations need to be compared in this range of
heliocentric distances, rather than at 1 AU from the Sun, to test
whether the results from the standard law and the modified law are
generally consistent.  This test shows that the radial components from
the two laws reach the same value, $+0.3 \times
10^{-5}$\,AU\,day$^{-2}$, at \mbox{$r = 0.639$ AU}, while the normal
components equally reach $+0.6 \times 10^{-5}$\,AU\,day$^{-2}$ at
\mbox{$r = 0.648$ AU}, in either case well inside the observed range of
heliocentric distances.  The comet's favorable comparison with the dwarf
sungrazers, mentioned in Section 6.2, is consistent with the results from
the modified-law solutions:\ C/1945 X1 appears to be subjected to a
nongravitational acceleration that is near the lower end of the range
of accelerations typical for the dwarf sungrazers.

In summary, we obtain some evidence suggesting that (i)~the nongravitational
solutions fit the observations of C/1945 X1 better than the{\vspace{-0.04cm}}
gravitational solutions; and (ii)~a law with a very steep slope (of the type
$\sim \!\! r^{-25}$) may be superior to the standard law.  Over all, however,
the promising performance of the modified law has not overturned our
conclusion expressed at the end of Section 6.2 that the limited data sample
does not make a conclusion about the existence of large nongravitational
effects in the motion of C/1945~X1 sufficiently compelling.  Every effort
should be expended to improve the comet's orbit by extending the observed
arc with the help of additional images on Boyden patrol plates.

\begin{table*}[t]
\begin{center}
\vspace{0.13cm}
{\footnotesize {\bf Table 11}\\[0.08cm]
Geocentric Ephemeris for Comet C/1945 X1 from 1945 September 12 to
1946 January 30 (Equinox J2000).\\[0.1cm]
\begin{tabular}{l@{\hspace{0.08cm}}c@{\hspace{0.06cm}}r@{\hspace{0.55cm}}c@{\hspace{0.25cm}}c@{\hspace{0.35cm}}c@{\hspace{0.4cm}}c@{\hspace{0.4cm}}c@{\hspace{0.15cm}}c@{\hspace{0.3cm}}c@{\hspace{0.3cm}}c@{\hspace{0.4cm}}c@{\hspace{0.3cm}}c@{\hspace{0.55cm}}c@{\hspace{0.3cm}}c@{\hspace{0.55cm}}c@{\hspace{0.3cm}}c}
\hline\hline\\[-0.22cm]
& & & \multicolumn{4}{@{\hspace{-0.28cm}}c}{From solution ${\cal N}_{\rm
 mod}^{\,(\!{\bf N}\!)}[\:\!0.05\!\:]\{\Im_0\}$} & &
 & \multicolumn{2}{@{\hspace{-0.2cm}}c}{App.\,magnitude}
 & \multicolumn{6}{@{\hspace{0cm}}c}{Differential ephemerides from
 solutions} \\[-0.05cm]
& & & \multicolumn{4}{@{\hspace{-0.28cm}}c}{\rule[0.6ex]{4.4cm}{0.4pt}} & Phase 
& & \multicolumn{2}{@{\hspace{-0.2cm}}c}{\rule[0.6ex]{2.1cm}{0.4pt}}
& \multicolumn{6}{@{\hspace{0cm}}c}{\rule[0.6ex]{6.3cm}{0.4pt}}\\[-0.05cm]
\multicolumn{3}{@{\hspace{-0.4cm}}c}{Date TT} & R.A. & Dec. & $\Delta$ & $r$
 & angle & PA({\boldmath $\!RV$}) & \,$r^{-4}$ & \,$r^{-7}$
 & \multicolumn{2}{@{\hspace{-0.2cm}}c}{${\cal N}_{\rm std}^{\,(\!{\bf
 N}\!)}\{\Im_0\}$} & \multicolumn{2}{@{\hspace{-0.15cm}}c}{${\cal G}\{\Im_0\}$}
 & \multicolumn{2}{@{\hspace{0.1cm}}c}{orbit D$^{\:\!\prime}$} \\[0.08cm]
\hline\\[-0.27cm]
1945 & Sept. & 12 & 7$^{^{\rm h}}\:\!\!\!$39$^{^{\!\rm m}}\!\!\!$.0
     & $\:\,-$10$^{^\circ}\!$00$^{^\prime}$ & 2.860 & 2.466
     & 20$^\circ\!\!$.1 & 258$^\circ\!\!$.5 & 15.7 & 19.7
     & $-$0$^{^{\rm S}}\!\!\!$.31 & +2$^{^{\prime\!\!\:\prime}}\!\!\!$.5
     & $-$2$^{^{\rm S}}\!\!\!$.8 & +117\rlap{$^{^{\prime\!\!\:\prime}}$}
     & $-$0$^{^{\rm S}}\!\!\!$.25 & +10$^{^{\prime\!\!\:\prime}}\!\!\!$.3 \\
   &      & 22 & 7 50.1 & $-$11 48 & 2.592 & 2.310 & 22.7 & 263.5
          & 15.2 & 19.0 & $-$0.31 & +2.7 & $-$2.8 & +116 & $-$0.26 & +10.3 \\
   & Oct. & $\;\:$2 & 8 01.6 & $-$14 03 & 2.315 & 2.148 & 25.6 & 268.2
          & 14.6 & 18.2 & $-$0.32 & +2.8 & $-$2.9 & +115 & $-$0.27 & +10.4 \\
   &      & 12 & 8 13.7 & $-$16 54 & 2.031 & 1.980 & 28.8 & 272.8
   & 14.0 & 17.3 & $-$0.33 & +3.0 & $-$3.0 & +113 & $-$0.29 & +10.4 \\[0.08cm]
   &      & 22 & 8 27.0 & $-$20 36 & 1.744 & 1.804 & 32.5 & 277.2
          & 13.3 & 16.3 & $-$0.34 & +3.2 & $-$3.1 & +110 & $-$0.31 & +10.4 \\
   & Nov. & $\;\:$1 & 8 42.3 & $-$25 33 & 1.458 & 1.619 & 37.2 & 281.5
          & 12.4 & 15.1 & $-$0.35 & +3.4 & $-$3.2 & +107 & $-$0.35 & +10.3 \\
   &      & 11 & 9 02.4 & $-$32 30 & 1.180 & 1.423 & 43.5 & 285.4
          & 11.4 & 13.6 & $-$0.38 & +3.5 & $-$3.3 & $\;\:$+94 & $-$0.40
          & +10.0 \\
   &      & 21 & 9 35.1 & $-$42 48 & 0.922 & 1.213 & 53.0 & 286.9
          & 10.2 & 11.9 & $-$0.42 & +3.2 & $-$3.5 & $\;\:$+71 & $-$0.53
          & $\;\:$+8.9 \\[0.08cm]
  & Dec.  & $\;\:$1 & \llap{1}0 54.6 & $-$57 49 & 0.713 & 0.982 & 69.1 & 277.1
          & $\;\:$8.7 & $\;\:$9.7 & $-$0.46 & +1.5 & $-$3.3 & $\;\:$+30
          & $-$0.90 & $\;\:$+5.6 \\
  &       & $\;\:$6 & \llap{1}2 34.0 & $-$65 04 & 0.645 & 0.855 & 80.7 & 256.5
          & $\;\:$7.9 & $\;\:$8.5 & $-$0.36 & +0.1 & $-$2.1 & $\;\:\;\:$+8
          & $-$1.22 & $\;\:$+1.7 \\
  &       & 11 & \llap{1}5 11.8 & $-$65 27 & 0.617 & 0.719 & 94.6 & 219.8
          & $\;\:$7.0 & $\;\:$7.0 & $-$0.03 & +0.1 & $-$0.1 & $\;\:\;\:$+1
          & $-$0.97 & $\;\:-$3.4 \\
  &       & 16 & \llap{1}7 12.4 & $-$56 11 & 0.640 & 0.569 & \llap{1}08.9
          & 189.8 & $\;\:$6.1 & $\;\:$5.3 & +0.04 & +1.7 & $-$0.3
          & $\;\:\;\:-$4 & $-$0.39 & $\;\:-$5.9 \\[0.08cm]
1946 & Jan.& $\;\:$5 & \llap{1}8 32.1 & $-$42 17 & 0.662 & 0.434 & \llap{1}26.4
          & 201.3 & $\;\:$5.0 & $\;\:$3.4 & +0.99 & $-$8.7 & +9.8 & +204
          & +1.10 & +19.9 \\
  &  & 10 & \llap{1}9 14.6 & $-$55 37 & 0.566 & 0.601 & \llap{1}14.8 & 184.7
          & $\;\:$6.1 & $\;\:$5.5 & +2.02 & $-$7.4 & $\!\!\!$+22.1 & +219
          & +2.12 & +22.3 \\
  &  & 20 & 0 37.9 & $-$69 21 & 0.525 & 0.882 & 84.7 & 104.0 & $\;\:$7.6
          & $\;\:$8.2 & +0.66 & $\!\!\!$+17.4 & $\!\!\!$+36.6 & +508 & +1.78
          & +44.0 \\
  &  & 30 & 3 32.8 & $-$48 43 & 0.664 & 1.124 & 60.6 & $\;\:$70.5 & $\;\:$9.1
          & 10.6 & $-$1.47 & +4.7 & $-$2.8 & +723 & $-$0.87 & +55.6 \\[0.07cm]
\hline\\[0cm]
\end{tabular}}
\end{center}
\end{table*}

\subsection{Astrometric and Orbital Accuracy}
It may seem to be questionable to use the astrometric data of less than high
quality for supporting the type of in-depth orbital analysis of C/1945~X1
that is described in the previous sections.  With the observed arc of merely
four days and the images taken with small cameras of poor spatial resolution
and long exposures tracked at sidereal rates, the data may at first sight
seem overexploited.  However, the addressed objectives are not only to learn
as much as possible about the orbital behavior of C/1945~X1 in the context
of other Kreutz sungrazers, but, as illustrated in Table 11, also to provide
a basis for constraining an ephemeris for the comet's motion both back and
forward in time from the observed period, a capability that is amply put to
use in our search for more images of the comet, as described in Section 7.

\section{Search for More Data on Comet C/1945 X1}
The preceding sections demonstrate that, in spite of some evidence to the
contrary, the current level of our knowledge of comet C/1945 X1 is less than
sufficient for addressing the issue of whether or not this object was indeed
a dwarf Kreutz sungrazer. It is essential that, where at all possible, existing
information be rechecked, reanalyzed, and reinterpreted, and that a vigorous
search for, and examination of, additional data be undertaken.  All our efforts
will be aided substantially by products of the Harvard College Observatory's
DASCH Project, currently in progress, that we referred to in Section 3.~As
for~the new data, our proposed work has three~\mbox{objectives}:\ (i)~to
identify, astrometrically measure, and reduce any further images of the comet
in the collection of digitized Boyden patrol plates (once they become
available) with the aim to refine the orbit by extending the arc covered by
observations as far back in time as possible; (ii)~to constrain the comet's
physical behavior by deriving magnitudes from all relevant digitized
photographs and compare the comet's light curve with that of other
sungrazers; and (iii) to search for and examine the comet's post-perihelion
images taken at Boyden for any traces of the object or its debris, including
the presence, appearance, and orientation of a headless tail, diagnostic of
the object's fate near perihelion.
\begin{table*}[t]
\begin{center}
\vspace{0.13cm}
{\footnotesize {\bf Table 12}\\[0.08cm]
Additional Boyden Plates with Possible Images of Comet C/1945 X1.\\[0.1cm]
\begin{tabular}{c@{\hspace{0.3cm}}l@{\hspace{0.07cm}}c@{\hspace{0.05cm}}r@{\hspace{0.5cm}}c@{\hspace{0.2cm}}c@{\hspace{0.2cm}}c@{\hspace{0.2cm}}c@{\hspace{0.5cm}}c@{\hspace{0.2cm}}c@{\hspace{0.25cm}}c@{\hspace{0.4cm}}c@{\hspace{0.4cm}}c@{\hspace{0.4cm}}c@{\hspace{0.2cm}}c}
\hline\hline\\[-0.22cm]
  & & & & \multicolumn{2}{@{\hspace{-0.1cm}}c}{Plate center$^{\rm b}$}
  & Exp. & & \multicolumn{6}{@{\hspace{-0.2cm}}c}{Comet's position
  (absolute and relative)$^{\rm b,c}$} & \\[-0.05cm]
Plate & \multicolumn{3}{@{\hspace{-0.3cm}}c}{UT time at}
 & \multicolumn{2}{@{\hspace{-0.1cm}}c}{\rule[0.6ex]{2.25cm}{0.4pt}} & time
 & Logbook & \multicolumn{6}{@{\hspace{-0.2cm}}c}{\rule[0.6ex]{6.4cm}{0.4pt}}
 & \\[-0.05cm]
number\rlap{$^{\rm a}$} & \multicolumn{3}{@{\hspace{-0.35cm}}c}{mid-exposure}
         & \,\,R.A. &  \,Dec. & (min) & reference & R.A. & Dec. & Plate
         & Dist. & P.A. & Edge\, & Observer\rlap{$^{\rm d}$}\\[0.06cm]
\hline\\[-0.22cm]
RB\,14049 & 1945 & Sept. & 19.11997
          & \,8$\;\!\!^{^{\rm h}}\!$04$\:\!\!^{^{\rm m}}\!\!\!\!\!\:$.6
          & $-$15$^\circ\!$17\rlap{$^\prime$} & 60 & rb10\_148
          & 7$^{^{\rm h}}\!$46$^{^{\rm m}}\!\!\!\!$.9
          & $\:\,-$11$\:\!\!^\circ\!\!\;$13$^\prime$ & ON & 5$^\circ\!\!$.91
          & 312$\!\!\:^\circ\!\!$.6 & 6$\!\!\;^\circ\!\!$.85 & Bl(?) \\
AM\,25110 & & Oct. & 10.07683 & 7 04.5 & $-$15 09 & 85 & am45b\_0144
       & 8 11.4 & $-$16 18 & ON & \llap{1}6.11 & $\;\:$96.3 & 1.18 & du Toit \\
RB\,14086 & &      & 17.07399 & 8 04.0 & $-$30 17 & \llap{1}20 & rb10\_132
          & 8 20.3 & $-$18 39 & ON & \llap{1}2.20 & $\;\:$18.5 & 2.33 & J(?) \\
AM\,25132 & &      & 27.06607 & 9 03.5 & $-$45 24 & 60 & am45b\_0148
          & 8 34.4 & $-$22 55 & OFF & \llap{2}3.25 & 342.8 & 0.71 & Bl(?) \\
AM\,25145 & & Nov. & 5.05881  & 9 04.2 & $-$30 24 & 90 & am45b\_0150
          & 8 49.7 & $-$28 04 & ON & 3.93 & 305.4 & \llap{1}3.99 & Bester \\
AM\,25150 & &      & 12.06740 & \llap{1}0 04.0 & $-$45 29 & 80 & am45b\_0152
          & 9 05.1 & $-$33 25 & ON & \llap{1}6.54 & 311.8 & 4.87 & J(?) \\
RB\,14121 & &      & 13.03559 & 8 03.2 & $-$45 17 & \llap{1}20 & rb10\_138
 & 9 07.6 & $-$34 16 & OFF & \llap{1}6.51 & $\;\:$53.8 & 2.12 & Bl(?) \\
AM\,25169 & &      & 15.06371 & \llap{1}0 33.7 & $-$60 31 & 85 & am45b\_0154
          & 9 13.4 & $-$36 11 & OFF & \llap{2}7.50 & 323.1 & 0.50 & J(?) \\
AM\,25187 & & Dec. & 7.04693  & \llap{1}1 04.6 & $-$45 32 & 90 & am45b\_0156
   & \llap{1}3 04.0 & $-$65 58 & OFF & \llap{2}5.94 & 152.4 & 1.49 & du Toit \\
AM\,25190 & &      & 8.01580  & \llap{1}0 33.7 & $-$60 31 & 90 & am45b\_0156
  & \llap{1}3 34.2 & $-$66 28 & ON & \llap{2}0.50 & 126.1 & 0.63 & J(?) \\
AM\,25208 & & & 14.03336 & \llap{1}6 40.9 & $-$70 12 & 40 & am45b\_0158
        & \llap{1}6 34.1 & $-$60 38 & ON & 9.60 & 355.0 & \llap{1}1.94 & J(?) \\
AM\,25231\rlap{$^{\rm e}$} & 1946 & Jan. & 5.04600 & \llap{1}2 05.2 & $-$70 33
          & 90 & am45b\_0162 & \llap{1}8 32.4 & $-$ 42 24 & OFF & \llap{5}2.65
          & 112.7 & \llap{3}1.37 & Bl(?) \\
RB\,14183\rlap{$^{\rm e}$} & & & 6.04985 & \llap{1}2 05.2 & $-$60 33
          & \llap{1}15 & rb10\_148 & \llap{1}8 38.1 & $-$44 57 & OFF
          & \llap{5}5.57 & 121.9 & \llap{3}5.99 & Bl(?) \\
RB\,14184\rlap{$^{\rm f}$} & & & 8.07729 & \llap{1}7 17.5 & $-$47 19 & 40
          & rb10\_148 & \llap{1}8 53.5 & $-$50 19 & OFF & \llap{1}6.02
          & 109.7 & 3.88 & Bester \\
AM\,25240 & &      & 21.79113 & 0 05.1 & $-$44 27 & 90 & am45b\_0164
          & 1 36.3 & $-$66 22 & OFF & \llap{2}5.13 & 158.5 & 1.89 & Britz \\
AM\,25241 & &      & 21.84446 & 1 33.6 & $-$59 29 & 60 & am45b\_0164
          & 1 37.8 & $-$66 15 & ON & 6.78 & 176.5 & \llap{1}4.73 & Britz \\
AM\,25244 & &      & 24.85635 & 5 25.4 & $-$68 55 & 90 & am45b\_0164
          & 2 40.0 & $-$59 43 & OFF & \llap{1}9.62 & 277.3 & 2.26 & Bl(?) \\
AM\,25260 & &      & 30.85589 & 5 03.9 & $-$29 52 & 90 & am45b\_0166
 & 3 38.4 & -47 05 & ON & \llap{2}3.86 & 217.8 & 2.56 & Bl(?) \\[0.05cm]
\hline\\[-0.25cm]
\multicolumn{15}{l}{\parbox{17.4cm}{$^{\rm a}$\,{\scriptsize AM plates
taken with the Cooke lens (covering 34$^\circ\!\!$.48 in R.A. and
43$^\circ\!\!$.10 in Dec.); RB plates with the Ross-Fecker camera
(22$^\circ\!\!$.32 by 27$^\circ\!\!$.90).}}}\\[-0.09cm]
\multicolumn{15}{l}{\parbox{14.8cm}{$^{\rm b}$\,{\scriptsize Equinox
J2000.}}}\\[0.03cm]
\multicolumn{15}{l}{\parbox{17.4cm}{$^{\rm c}$\,{\scriptsize Column
{\sf Plate} indicates whether or not the comet's predicted position
{\vspace{-0.06cm}}is within the plate limits; columns {\sf Dist.}\,and
{\sf P.A.\/} show the predicted position relative to the plate center;
and in column {\sf Edge} is the distance of the comet's image from the
plate edge (either on or off the plate).}}}\\[-0.08cm]
\multicolumn{15}{l}{\parbox{17.4cm}{$^{\rm d}$\,{\scriptsize The identities
of the observers whose abbreviations are Bl and J could not be determined,
major efforts notwithstanding; see also Table 1.}}}\\[-0.05cm]
\multicolumn{15}{l}{\parbox{17.4cm}{$^{\rm e}$\,{\scriptsize This
plate may show a segment of an early post-perihelion tail, if the comet
still produced significant amounts of dust (see Figure 6 or 7).}}}\\[-0.1cm]
\multicolumn{15}{l}{\parbox{17.4cm}{$^{\rm f}$\,{\scriptsize This is the
only deliberate attempt at detecting C/1945 X1 after perihelion; unfortunately,
the searched position is incorrect (see Figure 5).}}} \\[0.25cm]
\end{tabular}}
\end{center}
\end{table*}
\subsection{Search for Additional Images}
Our extensive orbit examination in Section 6 allows us now to extrapolate,
with a fair degree of confidence, the comet's motion way beyond the period of
December 11--15 and to realistically estimate uncertainties of an ephemeris
as a function of time.  The nominal geocentric ephemeris of the comet for the
period of time from 1945 September 12.0 TT to 1946 January 30.0 TT, presented
in Table 11, is computed{\vspace{-0.06cm}} from the nongravitational solution
${\cal N}_{\rm mod}^{\,(\!\bf N\!)}[0.05]\{\Im_0\}$ (whose elements
{\vspace{0.02cm}}and residuals are in Table 10), but{\vspace{-0.11cm}} the
differential ephemerides, derived from the solution ${\cal N}_{\rm std}^{(\!\bf
N\!)}\{\Im_0\}$ (elements and residuals in Table~9), from the gravitational
solution ${\cal G}\{\Im_0\}$ (Table~7), and from the orbit D$^\prime$ (see the
text near Table 6), are also included for comparison.  The two numbers listed
in Table 11 for each of these three additional solutions are the corrections
in right ascension and declination, respectively, that are to be added to the
nominal coordinates in columns~2 and~3; note the different units used.

The first six columns of Table 11 are self-explanatory.  The seventh column
gives the position angle of the prolonged radius vector (direction away
from the Sun), and the eighth and ninth columns include a predicted apparent
magnitude on two assumptions for the brightness variations:\ as an inverse
fourth and inverse seventh power of heliocentric distance, with no phase
effect.  Both versions were normalized to the comet's reported magnitude 7
at discovery on December 11, which predicts a magnitude of 9.5 at 1 AU from
both the Sun and the Earth for the inverse fourth-power law and 10.6 for the
inverse seventh-power law (see Section 7.2).

We find the agreement among the ephemerides based on the orbit D$^\prime$ and
the two nongravitational solutions quite remarkable.  On the other hand, the
ephemeris from the gravitational solution based on the same observations as
the two nongravitational orbits is very disappointing, differing from the
other three by as much as 0$^\circ\!$.2 over the period of 4.5 months.  We
note that the three consistent ephemerides come from the orbital sets that
imply that the line of apsides lies within 0$^\circ\!$.01 of the reference
value, while the ${\cal G}\{\Im_0\}$ set leaves an offset of nearly
0$^\circ\!$.08.

To further appraise the magnitude of ephemeris uncertainties, we completed
200 Monte Carlo runs with pure gravitational solutions fitted to the
$\{\Im_0\}$ set of observations, assuming, conservatively, measuring errors
of up to $\pm$9$^{\prime\prime}$ in the December 11--12 positions and up to
$\pm$2$^{\prime\prime}\!$.5 in the December 14--15 positions.  The standard
deviations from the 200 clones vary, in the 1945 September--November period,
between 2$^\prime\!$.4 and 3$^\prime$ in right ascension and between
1$^\prime\!$.4 and 5$^\prime$ in declination; in the 1946 January period,
the range is from 6$^\prime$ to nearly 20$^\prime$ in right ascension and
from 8$^\prime$ to 31$^\prime$ in declination.  On the average the offsets
of the ${\cal G}\{\Im_0\}$-based ephemeris in Table 11 amount to about
40\% of these standard deviations.  To sum it up, we believe that our nominal
ephemeris is accurate, in a worst case scenario, to a few arcmin over the
entire period of interest to us.

The next step is a search for all patrol plates whose fields cover (or could
cover) the comet's ephemeris position.  Ideally, the physical size of each
plate and its scale provide the angular distance from the center to an edge
as a function of a position angle.  It is in principle very easy to establish
whether the comet's image is or is not located within the plate's limits.
In reality, the problem is less straightforward because of a potentially
significant error in the reported position of the plate center.  Because of
this uncertainty (much larger than in the ephemeris), the comet's predicted
position may not be within the limits of a plate when it should, or vice
versa.  The search is further complicated by the comet's unknown brightness,
the observing conditions (the sky transparency, the moon interference, the
limiting magnitude, etc.), the exposure time, and the distribution of field
stars (needed for the astrometry), all of which determines whether the comet's
image, even if present on the plate, can in fact be recognized, astrometrically
measured and reduced, and photometrically evaluated.

Table 12 presents information on a total of 18 plates that may show an image
of C/1945 X1 or its debris, including the unreduced plate AM\,25208 from
December 14 (Table 1) but not the five from Table 4.  Because of the
position uncertainties, we include plates that nominally miss the comet's
position by up to 2$^\circ\!$.5.

These are all patrol plates, with very large fields of view (Section 3).
The Ross-Fecker plates (RB) cover a rectangle of 22$^\circ\!$.32 in right
ascension and 27$^\circ\!$.90 in declination ($\sim$600 square degrees),
the Cooke plates (AM) 34$^\circ\!$.48 by 43$^\circ\!$.10 ($\sim$1500 square
degrees).  The column of Table 12, titled {\sf Plate}, shows whether the
comet's predicted position, given by the standard polar coordinates relative
to the plate center, is (ON) or is not (OFF) within the limits of the plate.
The column {\sf Edge} indicates the distance of the comet's predicted
position from the plate edge.  The smaller the number in an ON case, the
more difficult it will be to identify and astrometrically evaluate the image
because of optical imperfections far from the optical axis of the instrument
and a potential lack of appropriate field stars.  The smaller the number in
an OFF case, the greater is a chance that the image could, after all,
appear on the plate because of the positional uncertainties.

We list a total of seven preperihelion plates on which the comet should show
up if bright enough and four such plates with the comet's predicted location
barely missed; see Section 7.2 for the brightness constraints.  After
perihelion there are only two potentially useful plates by the end of January
to check whether any material part of the head survived; two additional
plates just miss the comet's expected position.  On one of the plates taken
on January 21, we predict that the comet's location is not far from the
center.  The comet's head should also appear on five plates exposed during
February (AM\,25269 on the 7th, AM\,25272 and AM\,25273 on the 8th, RB\,14224
on the 19th, and RB\,14246 on the 27th), but there is no point in examining
these unless the comet shows up, rather unexpectedly, on the January plates.

Finally, there is a group of three plates (footnotes {\it e\/} and {\it f\/}
in Table 12), which miss the predicted position~of the comet's head by a wide
margin but may be relevant to a search for traces of the comet's dust debris
(Section 8.2).  This group includes the January 8 plate (Table~1), taken
with the Ross-Fecker camera specifically for the purpose to recover C/1945~X1
(Section 3).  Another plate, taken almost simultaneously with the Metcalf
refractor (Table~1), is not listed, because it is deemed useless on account
of its small field.

\subsection{Photometry and the Light Curve}
To predict the brightness of a comet is always risky, as has amply been
documented by historical examples.  For the Kreutz sungrazers, the task is
further complicated by apparent differences between the shapes of the
preperihelion light curves of a prominent Kreutz comet (observed as a bright
object from the ground) and a dwarf sungrazer (observed only from {\it
SOHO/STEREO\/}), as suggested by Knight et al.\ (2010).  However, a
preperihelion light curve is known for only one prominent sungrazer, C/1965~S1
(Sekanina 2002).\footnote{Only a single, crude preperihelion magnitude
estimate is available for C/1882 R1, the brightest Kreutz sungrazer over the
past two centuries.  This estimate is based on reports of its rivaling Venus
in brightness at the time of discovery, 12 days before perihelion (Gould
1883).}  It is not advisable to generalize this light curve as a standard
attribute of all bright Kreutz sungrazers.  This caveat is supported by
the perceived differences among the post-perihelion light curves for five
prominent Kreutz objects (C/1843 D1, C/1882~R1, C/1963~R1, C/1965~S1, and
C/1970 K1), which were fading with heliocentric distance at average rates
from $r^{-3.3}$ to $r^{-5.1}$ (Sekanina 2002).

Knight et al.\ (2010) concluded that the rate of brightening of the dwarf
Kreutz sungrazers observed with the {\it SOHO\/} coronagraphs changes
strikingly {\vspace{-0.035cm}}near 24 {\Rsun} (0.11 AU) from the Sun from
$\sim \! r^{-7.3}$ to $\sim \! r^{-3.8}$; this slower rate of brightening
then holds, according to them, down to 16 {\Rsun} (0.075 AU), where the
light curve begins to flatten.  They estimated that the steep rate is
unlikely to persist at heliocentric distances greater than 50 {\Rsun}
(0.23 AU) and that farther from the Sun the dwarf sungrazers brighten on
their way to perihelion by again following an $r^{-3.8}$ law.  However,
Ye et al.\ (2014) suggested that there is a significant diversity in the
preperihelion light curves of the dwarf Kreutz sungrazers.  They called
attention to either an outburst{\vspace{-0.04cm}} (whose amplitude exceeded
5~magnitudes) or an extremely steep brightening ($r^{-11}$ or steeper) of a
dwarf sungrazer C/2012~E2 between 1.06 AU and 0.52 AU from the Sun, while
the object followed, on the average, an $\sim\!r^{-4}$ (or flatter) law
during its approach to perihelion between at least 0.52 AU (rather than
0.23 AU) and 0.07 AU.  Ye et al.\ also pointed out that another dwarf
Kreutz sungrazer, C/2012~U3, was not detected at a heliocentric distance
of 1.31 AU before perihelion and was then much fainter than it should have
been if it followed the Knight et al.'s (2010) prediction.

Very illuminating is Ye et al.'s (2014) comparison of the light curves of
C/2012~U3 and C/2011~W3 (Lovejoy).  Although C/2011~W3 was not a dwarf
sungrazer, it brightened according to an $r^{-6.9}$ law (Sekanina \& Chodas
2012) from 0.75 AU down to 0.34 AU, thus behaving very differently from the
prominent sungrazer C/1965~S1 ($r^{-4.1}$ between 1.02 AU and 0.03 AU;
Sekanina 2002).  It appears that there is (i)~a distinct possibility that,
among the dwarf sungrazers, the steep, $r^{-7}$ law applies to heliocentric
distances of around 1 AU and possibly even further away from the Sun; and
(ii)~at least some of these objects appear to be subjected to outburst-like
brightening at moderate distances from the Sun on their way to perihelion.

The potential implications for the brightness evolution of C/1945 X1 are
obvious.  A major obstacle to accepting this object as a dwarf sungrazer
is its considerable brightness reported at the time of discovery.
Comparison with C/2011~W3 shows that C/1945~X1 was intrinsically (i.e.,
after a correction has been applied for the geocentric distance) fully
4.5~magnitudes\,(!) {\it brighter\/} at the same heliocentric distance.
Assuming the reported estimate is correct, this problem can only be overcome
if one accepts that C/1945 X1 experienced a major outburst some time before
the discovery, which was thereby substantially facilitated.  If the amplitude
of the outburst was greater than $\sim$4.5~magnitudes, and therefore comparable
to that of C/2012~E2, C/1945~X1 would have been fainter than C/2011~W3
before the onset of the outburst.  From the standpoint of chances that
C/1945~X1 could be detected on any of the pre-discovery plates in Table 12,
the pivotal parameter is the time when the putative event took place.  If
this general scenario is correct, the comet's unusual brightness is indeed
explained, but there is rather a slim chance of detecting the comet on
any of the pre-outburst plates. 

An alternative explanation of the reported magnitude of C/1945 X1, if it
should be a dwarf sungrazer, is simply a gross overestimation of its
brightness by the discoverer.  In any case, these considerations show the
critical need for a thorough photometric examination of the comet's Boyden
images, both those from the period of December 11--15 and any pre-discovery
ones.

Keeping all our options open, we illustrate the dramatic differences, over
a wide range of heliocentric distances, between the C/1945~X1 brightness
predictions based on the $r^{-4}$ and $r^{-7}$ laws by listing the apparent
magnitudes in columns 8--9 of Table 11.  One should, however, be able to
discriminate between the two laws over a much shorter range of $r$.  In
fact, in the known Boyden images taken over the period of December 11--15,
the comet should have brightened by 0.7~magnitude if the $r^{-4}$ law
applied, but by 1.3~magnitudes if the steeper law was in effect.  The
ephemeris suggests that C/1945 X1 might have been as bright as magnitude 15
in the second half of September, about 3 months before perihelion.  In order
that no image be missed, we began our search for plates that fit the predicted
position of the comet as early as mid-September, since the Ross-Fecker camera,
the more powerful of the two patrol instruments, has nominally a limiting
magnitude 15 as well (Section~3).

\section{What Kind of a Sungrazer Was C/1945 X1?}
%
%
Evidence of the fate of C/1945 X1 after its passing through perihelion
is vital for determining the place of this object in the hierarchy and
evolution of the Kreutz system and for answering the question of whether
it was a dwarf sungrazer.  We propose to establish this from comparison
with some better investigated sungrazers.

\subsection{C/1945 X1 and the Transition Sungrazers}
Based on large numbers of observations, the members of the Kreutz system
are usually divided into two categories:\ prominent (bright and often quite
spectacular as seen from the ground) and dwarf (defined in Section~1).
This is a very simplistic classification, because it ignores intermediate
objects, which occupy the transition between both categories and, albeit
scarce, are momentous for a better understanding of the disintegration
process of the Kreutz sungrazers.

In an effort to determine where C/1945 X1 is likely to fit in, we selected
three ``standards'' to define a scale for transition objects in the order of
increasing similarity to the dwarf members.  The standards were chosen to be
represented by C/2011~W3, C/1887~B1, and C/2007~L3.  Next we describe their
diagnostic properties, which will be searched for on the best suited plates
of C/1945~X1 in order to classify the comet on this scale.

{\bf C/2011 W3} was the most persistent and nearest~the status of a prominent
member among the three~\mbox{objects}.  With a perihelion distance of
1.19\,{\Rsun}, this comet survived intact (without its nucleus disintegrating
completely) until \mbox{$38 \pm 5$}~hours after perihelion, at which time
it suddenly fell apart in a {\it terminal outburst\/}, losing suddenly its
residual, $\sim$10$^{12}$\,g nucleus and head (Sekanina \& Chodas 2012).  The
tail, made up of microscopic dust particles ejected from the nucleus {\it
before\/} perihelion, was seen to survive perihelion for at least 24~hours
as a {\it syndyname\/} --- a locus of dust released at different times but
subjected to the same radiation pressure acceleration $\beta$ --- of
\mbox{$\beta = 0.6$} the solar gravitational acceleration, typical for
dielectric submicron-sized particles.  Only grains ejected less than 0.1~day
before perihelion, which approached the Sun within 1.8\,{\Rsun}, sublimated
away.  After perihelion the comet formed a new, spectacular, headless tail,
most of which was a product of the event $\sim$38 hours after perihelion.
This tail was a {\it synchrone\/} --- a locus of dust grains of different
sizes (subjected to a broad range of radiation pressure accelerations), all
released at the same time --- that was under observation for up to 90 days,
until mid-March 2012.  Its representative length during most reported sightings
again suggested a maximum radiation pressure acceleration of \mbox{$\beta
\simeq 0.6$} the solar gravitational acceleration.  Only during the late
observations (60--90 days after perihelion) did the visible tail consist of
merely micron-sized and larger grains, whose \mbox{$\beta \ll 0.6$} the solar
gravitational acceleration.

{\bf C/1887 B1} is the runner-up to C/2011 W3.  Available information is
rather limited, because this object was discovered only after perihelion as
a headless, narrow, ribbon-like tail, again a synchrone.  Nonetheless, the systematic
variations in this tail's orientation with time suggest that the nucleus
survived intact until \mbox{$5.8 \pm 0.8$} hours after perihelion (Sekanina
1984), at which point it must have suddenly disintegrated in a terminal
outburst similar to that of C/2011~W3.  At perihelion the comet was merely
1.04\,{\Rsun} from the Sun (Sekanina 1978; Seka\-nina \& Chodas 2004).  The
tail, truly spectacular during the early sightings from 8.5 days after
perihelion on, was under observation until 18.5~days after perihelion and
its length implied a maximum radiation pressure acceleration of dust particles
that dropped from $\sim$0.4 the solar gravitational acceleration at discovery
to $\sim$0.05 during the last three reported observations.  The minimum
particle size in the visible part of the tail was thus increasing with time
from less than 1~micron to several microns.

{\bf C/2007 L3} is, of the three standards, the most remote from the prominent
Kreutz-system members, being in fact a representative of a relative small group
of bright dwarf Kreutz sungrazers, which develop long, extremely narrow tails
upon their approach to perihelion and which we refer to hereafter as the {\it
superdwarfs\/}.  Like ordinary dwarf sungrazers, their nuclei fail to survive
perihelion, so that they form no post-perihelion tails. We chose C/2007~L3 as
a representative primarily because it was imaged by both {\it SOHO\/} and {\it
STEREO\/}, unlike most of its peers.  The comet had a perihelion distance of
1.530\,{\Rsun} (Marsden 2008), very close to our best estimate for C/1945~X1,
its brightness peaked at magnitude $\sim$3 and its head disappeared about
80~minutes before perihelion, at a heliocentric distance of 2.75\,{\Rsun}
(Green 2007).  The {\it SOHO\/} and {\it STEREO\/} images were studied in
considerable detail by Thompson (2009), who closely confirmed the earlier
conclusions by Sekanina (2000), based on several objects, that preperihelion
images of these tails show that they are synchrones, referring to
emission-shutoff times of \mbox{$26 \pm 7$} hours before perihelion
and to heliocentric distances at shutoff of \mbox{$23.5 \pm
4.5$\,{\Rsun}}.\footnote{For the best two imaged comets (C/1996~Y1 and
C/1998~K10) in his set, Sekanina (2000) noticed that the tail was
actually bent and, unlike its bright synchronic portion, the faint, outer
part was a syndyname with a range of the ejection times.{\vspace{0.05cm}}}
This average agrees remarkably well with Knight et al.'s (2010) distance at
which the slope of the light curves suddenly changes from $-$7.3 to $-$3.8
(Section 7.2).\footnote{This evidence may be interpreted to indicate that dust
emission essentially ceases at 24\,{\Rssun} from the Sun and the $r^{-3.8}$
law of brightening measures an effect of the object's fragmentation.}  The tail
of C/2007~L3 survived the comet's perihelion, as did the preperihelion tail of
C/2011~W3.  Thompson examined its remnant and found, unlike Sekanina \& Chodas
(2012) for C/2011~W3, that the feature remained synchronic to its last examined
image, nearly 17~hours after perihelion, but that its position was slightly
modified by a loss of angular momentum, possibly due to atmospheric drag from
the solar corona.  The apparent discrepancy is explained by the fact that the
dust emission of C/2011~W3 did not shut off about 1 day before perihelion, but
continued.  Both Sekanina (2000) and Thompson (2009) detected the maximum
radiation pressure accelerations of \mbox{$\beta \simeq 0.6$} the solar
gravitational attraction.  Comet C/2007~L3 was not the only Kreutz superdwarf
whose dust tail was observed to survive perihelion.  Other examples are
C/1979~Q1, the first {\it Solwind\/} comet (Michels et al.\ 1982), C/1998~K10,
one of the dozen comets examined by Sekanina (2000), etc.  It is likely that
most, if not all, superdwarfs would show their tails surviving perihelion, if
their images are closely analyzed.  About 10 such comets from 1995--2005 in
Knight et al.'s (2010) Table~2 and all 19 sungrazers from 2006--2013 in
Sekanina \& Kracht's (2013) Table~1 belong to this subcategory, which is at
present estimated to include about three dozen objects.

%
So where does C/1945 X1 fit in?  Employing the constraint that no tail was
ever reported with the unaided eye after perihelion, it is certain that this
comet was too faint to be placed between C/2011~W3 and C/1887~B1 or even near
C/1887~B1.  Since a {\it preperihelion\/} tail is always found to survive for
only 1 day or so after perihelion, there is no chance of detecting it on any of
the plates listed in Table 12.  Should we find traces of the comet's synchronic
tail, a product of a {\it post-perihelion\/} outburst, on any of these plates,
then C/1945~X1 is to be classified on this three-point scale between C/1887~B1
and C/2007~L3, in which case it was a transition object more massive than a
superdwarf.  If not, it has to be concluded that C/1945 X1 cannot be positioned
higher than C/2007~L3, so that, at best, it was a superdwarf.

\subsection{Search for Traces of C/1945 X1 After Perihelion}
\vspace{-0.04cm}
From these considerations it follows that while a search for, and examination
of, the comet's preperihelion images on the plates listed in Table 12 should
serve to improve our understanding of the comet's orbital motion and its
light curve, a careful search for traces of the comet on, and examination of,
the plates taken after perihelion should provide information on the comet's
fate and help answer the question of whether C/1945~X1 was a transition
object or a dwarf sungrazer; because of the lack of information on the
comet's appearance in close proximity of perihelion (Table 12), we cannot
distinguish between its classification as a dwarf or a superdwarf.

Before we discuss specific search opportunities, we~note that the dust released
from this comet before perihelion was affected by sublimation less severely than
the dust from C/2011~W3.  The orbital elements for C/1945~X1's preperihelion
ejecta subjected to radiation pressure accelerations \mbox{$\beta = 0.6$}
the solar gravitational acceleration are summarized in Table~13.  It follows
that if this comet's dust had sublimation properties similar to those of the
dust from C/2011~W3, only particles ejected within $\sim$1~hour of perihelion
should have sublimated completely, because of a greater perihelion distance of
C/1945~X1.  A major delay of several days in passing through perihelion is
noted as a yet another effect:\ for example, particles ejected 50 days before
the comet's perihelion did not reach their own perihelion until late on
January 5.

As to the sources of information on comet C/1945~X1 after perihelion, a remote
opportunity occurred on 1946 January 3.51 UT, the time of a solar
eclipse.\footnote{See the website {\tt http://eclipse.gsfc.nasa.gov/SEplot1901/ SE1946Jan03P.GIF}.} Unfortunately, it was merely partial, of magnitude~0.553,
and visible only from Antarctica and surrounding oceanic regions.  Even less of
the Sun was occulted at the locations of several permanent British scientific
stations in Antarctica near the southernmost tip of South America (Operation
Tabarin; e.g., Haddelsey 2014).  At J.\,Shanklin's suggestion we checked the
UK Met Office Archives in Exeter and, not surprisingly, found no report on
any relevant observations at the time.

Back at Boyden, an attempt was made by M.\,J.\,Bester on January~8 to recover
the comet; the search, as pointed out (Sections 3 and 7.1), was conducted with
the Metcalf refractor and the Ross-Fecker piggyback camera.  Both exposures
pointed in the same direction because they overlapped each other; the Metcalf
45 minute exposure began 20 minutes earlier and was longer by 5 minutes.  The
pointing of the telescope is, however, a mystery:\ as performed, it could not
accommodate the comet's predicted position within the limits of either plate.

\begin{table}[t]
\begin{center}
\vspace{0.15cm}
{\footnotesize {\bf Table 13}\\[0.1cm]
Orbital Elements of Dust Particles Subjected to $\beta = 0.6$\\ and Ejected
from Comet 1945 X1 up to Perihelion.\\[0.1cm]
\begin{tabular}{c@{\hspace{0.8cm}}c@{\hspace{0.75cm}}c@{\hspace{0.5cm}}c}
\hline\hline\\[-0.25cm]
$\;\;$Time of & Time of & Perihelion & Orbital \\[-0.03cm]
$\;\;$ejection & perihelion & distance & eccentricity\\[-0.02cm]
$\;\;$(days)\rlap{$^{\rm a}$} & (days)\rlap{$^{\rm a}$} & ({\Rssun}) & \\[.05cm]
\hline\\[-0.2cm]
$-$50     & +8.868 & 3.757 & 1.035 \\
$-$40     & +7.142 & 3.749 & 1.041 \\
$-$30     & +5.410 & 3.732 & 1.049 \\
$-$25     & +4.540 & 3.721 & 1.055 \\
$-$20     & +3.667 & 3.706 & 1.064 \\
$-$15     & +2.788 & 3.682 & 1.077 \\
$-$10     & +1.899 & 3.641 & 1.101 \\
$\;\:-$7  & +1.357 & 3.596 & 1.127 \\
$\;\:-$5  & +0.989 & 3.545 & 1.157 \\
$\;\:-$3  & +0.610 & 3.450 & 1.217 \\
$\;\:-$2  & +0.412 & 3.356 & 1.279 \\
$\;\:-$1  & +0.203 & 3.134 & 1.426 \\
$\;\:-$0\rlap{.7} & +0.136 & 3.027 & 1.527 \\
$\;\:-$0\rlap{.5} & +0.089 & 2.896 & 1.642 \\
$\;\:-$0\rlap{.3} & +0.041 & 2.673 & 1.861 \\
$\;\:-$0\rlap{.2} & +0.017 & 2.481 & 2.082 \\
$\;\:-$0\rlap{.1} & $-$0.004 & 2.144 & 2.567 \\
$\;\:-$0\rlap{.05} & $-$0.010 & 1.843 & 3.150 \\
$\;\;-$1\rlap{$^{\:\!\!\rm h}$} & $-$0.009 & 1.779 & 3.299 \\
$\;\;\;\;\,$0\rlap{.0} & $\;\;\,$0.000 & 1.530 & 4.000 \\[0.04cm]
\hline\\[-0.28cm]
\multicolumn{4}{l}{\parbox{7.2cm}{$^{\rm a}$\,{\scriptsize Reckoned from
the time of the comet's perihelion passage.}}}\\[-0.02cm]
\end{tabular}}
\end{center}
\end{table}

January 8 was the date of publication of Cunningham's (1946b) ephemeris, but
early in the morning Bester could not have been in the possession of the
official announcement from Copenhagen.  Yet, the telescope was set to point
precisely in accord with the ephemeris in declination and was --- most
peculiarly --- off by exactly 2~hours to the west of the ephemeris place
in right ascension.  Such a coincidence surely cannot be accidental.  We
presume that Bester was notified directly by Cunningham ahead of the ephemeris'
publication, perhaps by cable, but there was a miscommunication, an error in
one coordinate.  In any event, the time of Bester's observation was
fundamentally incorrect because during the exposures the comet, less than
30$^\circ$ from the Sun, was below and, later, very close to the horizon.
Since the comet was almost exactly south of the Sun, it could have been
observed either shortly after sunset or, somewhat preferably, just before
sunrise.  The need to take a fairly long exposure did of course constrain
the choice of the observation time.  If observed at the correct time with
the telescope set to point in accord with Cunningham's ephemeris in both
coordinates, the comet's position would have been exposed on both plates,
even though only $\sim$0$^\circ\!$.5 from the edge of the Metcalf plate.

\begin{figure*}[t]
\vspace{-3.4cm}
\hspace{0.2cm}
\centerline{
\scalebox{0.77}{
\includegraphics{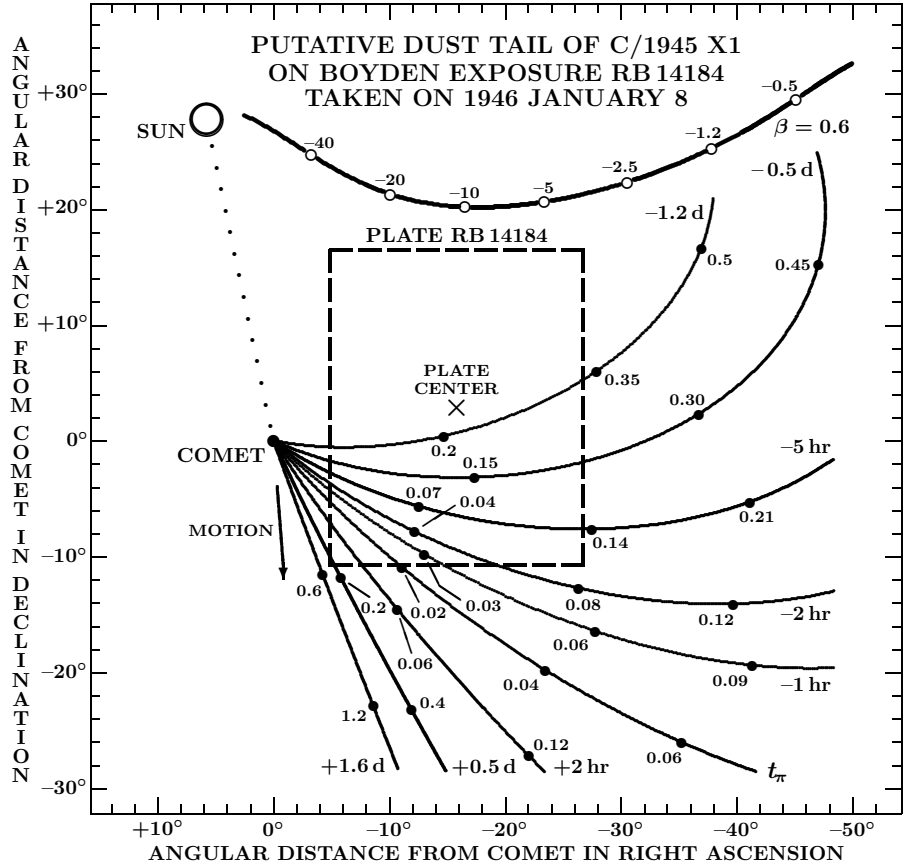}}} 
\vspace{-8.1cm}
\caption{The predicted projected position of C/1945 X1 relative to the sky
area exposed on the plate RB\,14814, taken by M.\,J.\,Bester with the
Ross-Fecker camera on 1946 January 8.07729 UT.  While the comet's head was
more than 4$^\circ$ off the eastern edge of the plate, much of the dust tail
made up of particles released near and shortly before perihelion occupy the
lower, southern part of the plate's field.  Dust ejected as far back as
$\sim$40 days before perihelion occupies the rest of the plate.  The curves to
the south of the plate center are the synchrones, the loci of the dust that was
ejected from the comet at particular times between 1.2 days before perihelion
and 1.6 days after perihelion; the curve $t_\pi$ is the perihelion synchrone.
The numbers along the synchrones are the values $\beta$, the ratio between
the solar radiation pressure acceleration and the solar gravitational
acceleration; $\beta$ and the ejection time govern the motions of the dust
ejecta.  Values near \mbox{$\beta \approx 0.1$} are typical for micron-sized
grains, those near or in excess of 0.5 refer to a variety of submicron-sized
grains.  The thick curve at the top, labeled \mbox{$\beta = 0.6$} is a
syndyname defined by this $\beta$ ratio and typical for dielectric
submicron-sized particles.  Shown along this curve are the times of ejection
in days before perihelion; see text for the significance of this
syndyname.{\vspace{0.55cm}}}
\end{figure*}

Figure 5 displays the projected positions of the comet and the Sun relative
to the pointing direction of the plate RB\,14184, taken with the Ross-Fecker
camera on January 8, 11 days after perihelion.  Although dust ejecta are
predicted to be all over the plate, their detection, contingent on the
time-dependent emission rate, is problematic on account of their generally
low surface brightness, except when released during a sharply peaked event,
such as a terminal outburst.  The curves emanating in the figure from the
comet's head and distributed mostly in the lower part of the plate are the
near-perihelion synchrones.  Their selected range is from 1.2 days before
perihelion to 1.6 days after perihelion.  Few post-perihelion ejecta project
onto the plate, barely touching its southeastern corner, and only a limited
fraction of the perihelion ejecta crosses the plate.  On the other hand, much
of the dust released as far back as  $\sim$40~days before perihelion is
projected onto the plate.  We emphasized in Section 8.1 that the preperihelion
tails of the transition object C/2011~W3 and of C/2007~L3 and many other (if
not all) superdwarfs survived perihelion for not much longer than $\sim$1~day
at the most.  Such features are too close to the Sun to observe from the
ground.  It is only for the sake of interest that Figure~5 shows the syndyname
\mbox{$\beta = 0.6$}, made up of dielectric submicron-sized particles ejected
from C/1945~X1 before perihelion, to miss the field of the sky covered by the
plate RB\,14184, nearly grazing it along the entire northern edge.

The plate RB\,14184 may prove useful.  If C/1945~X1 experienced a terminal
outburst and disintegrated within an hour or two after perihelion and if the
ejecta contained a large abundance of grains several microns in diameter,
there is a possibility that this debris could show up as a band of light
across the lower left corner of the plate.  The phase angles favor a modest
forward-scattering effect, being mostly in a range of 100$^\circ$--110$^\circ$.
Submicron-sized ejecta released from the comet at the same times should miss
the plate.  Thus, a  detection or nondetection of a post-perihelion synchronic
tail could provide meaningful constraints on the existence of a terminal
outburst and on the time of nucleus disintegration.

We found two more plates that might provide additional limited information
on a potential post-perihelion terminal outburst:\ AM\,25231, taken with
the Cooke lense in the morning of January~5, and RB\,14183, taken with
the Ross-Fecker camera 24~hours later.  They both are listed in Table 12.
The synchrones for the January~5 plate are depicted in Figure~6, those
for the January 6 plate in Figure~7.  The positions of dust particles
relative to the comet's nucleus, determined by a computer code in terms of
their polar coordinates, were converted to~the polar coordinates of the
offsets from the plate center, using a technique described in some detail in
the Appendix.  An advantage of these plates relative to that on RB\,14184 is
their exposure times, 115 and 90~minutes, respectively, as opposed to only
40~minutes (Table~12).  The January~5 plate might provide useful information
only if the terminal outburst took place between, at most, $\sim$2 and
$\sim$4~hours after perihelion; a synchronic tail containing earlier ejecta
would miss this plate, while segments of later synchrones limited to
\mbox{$\beta > 0.6$} the solar gravitational acceleration are unlikely to
refer to any real tails (Section 8.1).  Similarly, the January~6 plate might
provide information only on an outburst that occurred between $\sim$1 and
$\sim$5~hours after perihelion.  No forward-scattering effect can be expected
in either case, as the phase angles range between about 60$^\circ$ and
80$^\circ$.

\begin{figure}[t]
\vspace{-4.69cm}
\hspace{2.45cm}
\centerline{
\scalebox{0.68}{
\includegraphics{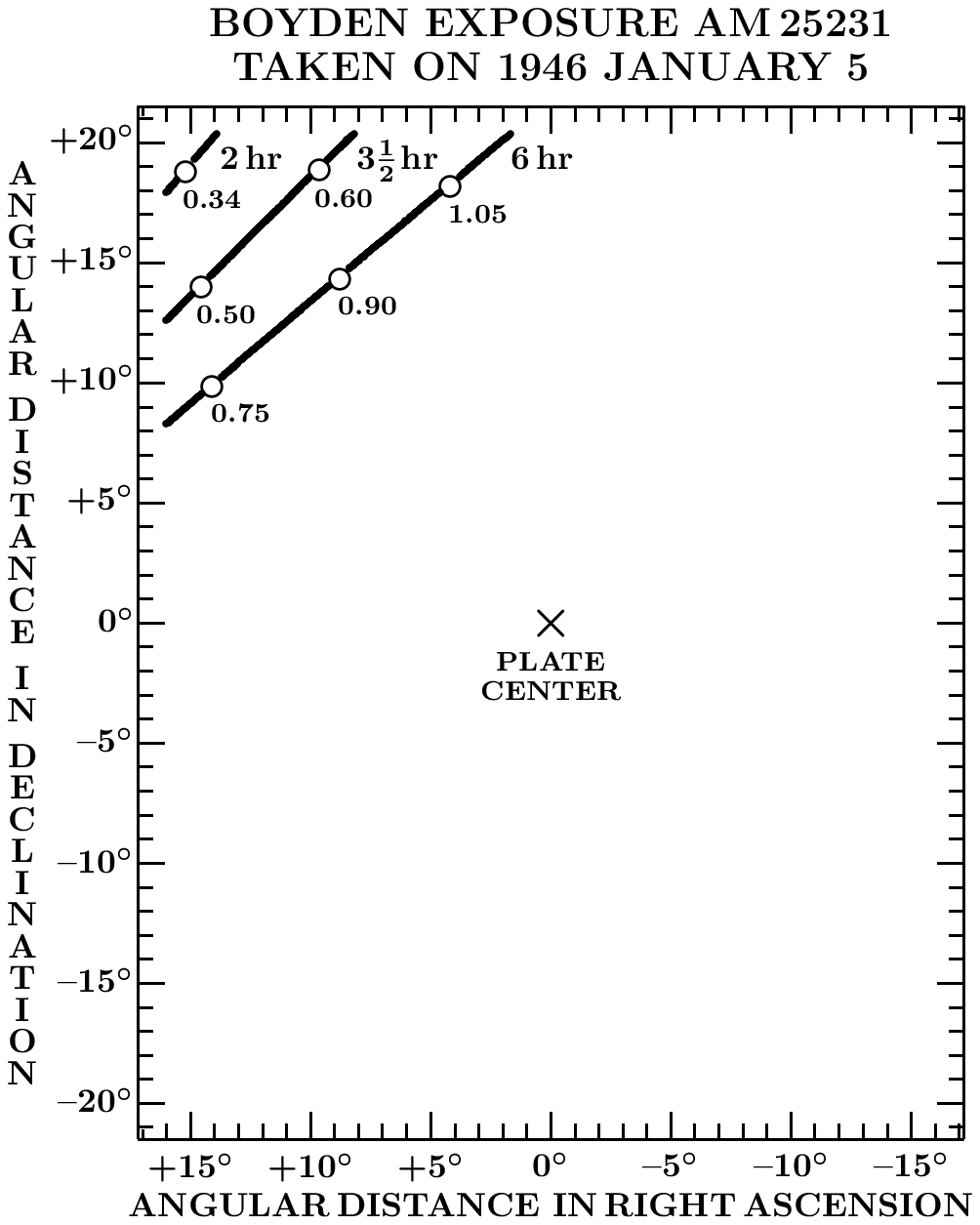}}} 
\vspace{-7.15cm}
\caption{Predicted positions, on a Cooke plate AM\,25231 taken on
January 5.04600 UT, of synchrones that depict C/1945~X1's dust ejecta
from hypothetical terminal outbursts 2, 3.5, and 6 hours after perihelion.
The smaller-font digits refer to particles subjected to a variety of radiation
pressure accelerations $\beta$ between 0.34 and 1.05 the solar gravitational
acceleration. Since the Kreutz comets appear to have \mbox{$\beta \leq
0.6$}, only outbursts that occurred between $\sim$2 and $\sim$4~hours after
perihelion might be detected on this plate.{\vspace{0.45cm}}} 
\end{figure}

In the highly unlikely case that a major fragment of the original nucleus
survived for weeks after perihelion, the best search opportunity is offered by
a Cooke plate AM\,25241, taken on January 21, on which the comet is predicted
to be less than 7$^\circ$ from the center (Table 12).  Less favorable
circumstances should accompany a search on a plate AM\,25260, taken 9~days
later, on which the comet's predicted position is only 2$^\circ\!$.5 from
the edge.  Additional search opportunities are more discouraging~still.

\section{Conclusions}
Comet C/1945 X1 has been rather an oddball among the Kreutz sungrazers observed
from the ground ever since its discovery 70 years ago.  Although the dilatory
handling of the Boyden plates contributed to the snail pace in getting out the
facts about the object, it was nonetheless its apparent lack of luster near
and after perihelion that is primarily responsible for our ignorance of its
properties, fate, and place in the Kreutz system.

\mbox{The$\:$comet's$\:$disappointing$\:$post-perihelion\,performance} ---
\mbox{especially$\:$the$\:$absence$\:$of$\:$a$\:$prominent$\:$dust$\:$tail} ---
pro\-vides a strong argument against
its being on a par with the headless sungrazers C/2011~W3 and C/1887~B1 that
survived perihelion with a fairly massive nucleus intact but disintegrated
shortly afterwards.~The~\mbox{plausible}~scenarios for C/1945~X1 are
thus limited to~two:\ either~the comet's surviving mass was sufficient to
generate a modest,\,but not spectacular,\,post-perihelion tail,\,or~the~comet
completely disintegrated still before perihelion. While there is not much
of a difference between~the~two~scenarios,\,\mbox{we use the second one to
define a dwarf Kreutz}~sun\-grazer.  The existence of a terminal outburst
is therefore critical to distinguish between the two options.

\begin{figure}[t]
\vspace{-4.33cm}
\hspace{2.35cm}
\centerline{
\scalebox{0.658}{  
\includegraphics{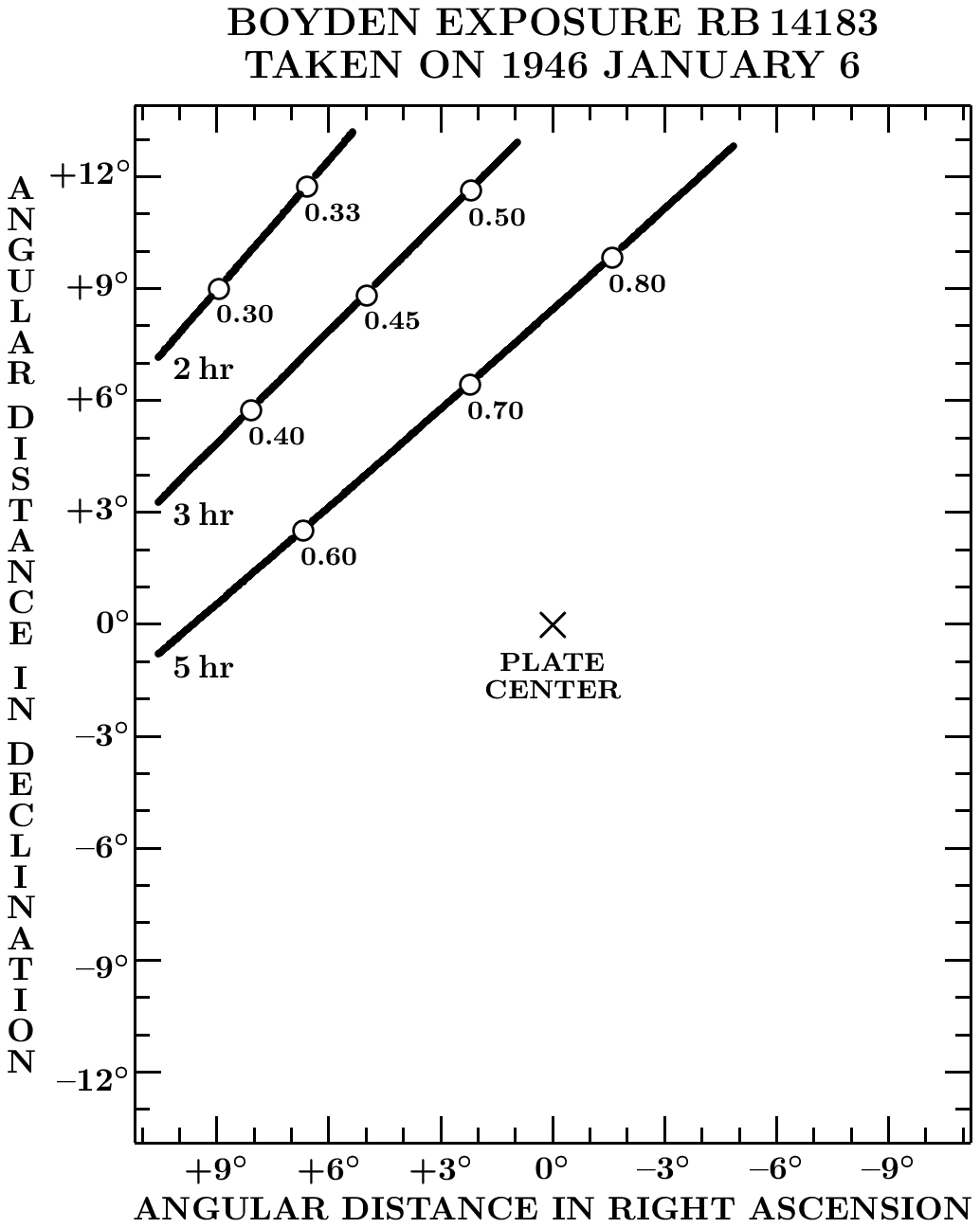}}} 
\vspace{-6.86cm}
\caption{Predicted positions, on a Ross-Fecker plate RB\,14183 taken on
January 6.04985 UT, of synchrones depicting the comet's dust ejecta from
hypothetical terminal outbursts 2, 3, and 5~hours after perihelion.  The
smaller-font digits refer to particles subjected to a variety of radiation
pressure accelerations $\beta$ between 0.3 and 0.8 the solar gravitational
acceleration.  Since the Kreutz comets appear to have \mbox{$\beta \leq
0.6$}, only outbursts that occurred between $\sim$1 and $\sim$5 hours after
perihelion might be detected on this plate.{\vspace{0.3cm}}}
\end{figure}

Taking account of the indirect planetary perturbations on the premise that
a precursor of C/1945 X1 separated from its common parent with C/1882 R1
(\mbox{Sekanina} \& Chodas 2004), we determine that the line of apsides of
C/1945 X1 should point toward an ecliptical longitude of \mbox{$L_\pi =
282^\circ\!.44$} and latitude of \mbox{$B_\pi = +35^\circ\!.16$} (Equinox
J2000) and that the comet's osculating semimajor axis $a$ should be 96.70 AU
at an epoch 1945 December 28.0 TT or 96.67 AU at an epoch December 11.0 TT.

Employing this value of $a$ as an orbital constraint, our {\it Best Fit\/}
purely gravitational solution based on all five observations left an
unacceptably large offset of 0$^\circ\!$.57 from the nominal apsidal-line
direction, apparently because the third observation was inferior.  This was
confirmed by four-observation gravitational solutions, of which the one
based on the positions from December 11, 12, 14, and 15 was by far the
most promising, leaving an offset from the nominal apsidal-line direction
of 0$^\circ\!$.08, much better but still not entirely satisfactory.

Applying a technique introduced recently by us in Paper 1, we show that the
incorporation of a nongravitational acceleration of the orbital motion
improves the fit in terms of both the apsidal-line offset and mean residual.
Use of Marsden et{\vspace{-0.04cm}} al.'s (1973) standard formalism suggests
an acceleration on the order of 10$^{-6}$\,AU day$^{-2}$ (with a relative
error of $\pm$8--17\%), near the lower end of a range typical for the dwarf
sungrazers.  A preferable nongravitational model, based on a much steeper
acceleration law (apparently common among the dwarf sungrazers; see Paper~1)
than is the standard law, leads to the same magnitude of the integrated
effect but fits the observations better, with the relative errors of the
nongravitational parameters reduced to $\pm$5--7\%.

In order to augment the current, very limited data set, an extensive search
for further possible images of the comet should be undertaken.  We hope
to be able to conduct such a search, based on our orbital computations
suggesting that at least 7 and perhaps as many as 11 preperihelion Boyden
patrol plates taken between late September and mid-December may contain
such images.  The plates will be inspected in the near future, once their
digitized copies are made available by the DASCH Project.  At present we
do not anticipate to expand our search to archives of wide-field plates
from other observatories (e.g., Tsvetkov \& Tsvetkova 2012).

In addition to positional data of the comet, its photometry will also have
to be performed to reexamine the reported brightness at discovery and to
settle the issues of the rate of brightness variations with heliocentric
distance and the chance of an early preperihelion outburst. Comparison with
C/2011~W3 and C/2012~E2, among others, may help settle some of the issues of
preperihelion activity of the Kreutz sungrazers.

Finally, to pursue a solution to the complex problem of C/1945 X1's place in
the hierarchy of the Kreutz system and the comet's fate, we also will
examine~some post-perihelion Boyden patrol plates to search for both potential
relics of the comet itself and/or possible traces of its dust tail as a product
of a modest terminal outburst that may have occurred just hours past perihelion.
Whereas it appears that, masswise, C/1945~X1 could~not compete with either
C/2011~W3 or C/1887~B1, we will try to disentangle a mystery of whether it
still could be considered a transition object distinctly more massive than
a superdwarf, such as C/2007~L3, C/1998~K10, or C/1979~Q1, or it does not
differ materially from the Kreutz dwarfs or superdwarfs --- and thus answer
the cardinal question posed in this paper's title.\\[-0.2cm]

The authors thank D.\ van Jaarsveldt, J.\ Grindlay, A.\ Doane, and E.\ Los
for their information on the observers and the plate collection from the
Boyden Station and on the Harvard College Observatory's DASCH Project.
This research was carried out in part at the Jet Propulsion Laboratory,
California Institute of Technology, under contract with the National
Aeronautics and Space Administration.\\[-0.27cm]
\begin{center}
{\footnotesize APPENDIX \\[0.2cm]
TRANSFORMATION OF DUST PARTICLE POSITION FROM \\
COMETOCENTRIC TO PLATE CENTER COORDINATES} \\
\end{center}
The code computing the position of a dust particle, which is a function of
the ejection time, $t_{\rm eject}$, and a ratio of the radiation pressure
acceleration to the solar gravitational acceleration, $\beta$, provides the
polar coordinates in reference to the comet's nucleus in projection onto the
plane of the sky --- the angular distance from the nucleus, $D_{\aster}(t_{\rm
eject},\beta)$, and the position angle, $\Pi_{\aster}(t_{\rm eject},\beta)$,
reckoned from the north through the east.  The equatorial coordinates of the
nucleus, right ascension $\alpha_0$ and declination $\delta_0$, and the
coordinates of a particle, $\alpha_{\aster}(t_{\rm eject},\beta)$ and
$\delta_{\aster}(t_{\rm eject}, \beta)$, are related to $D_{\aster}$ and
$\Pi_{\aster}$ by the well-known expressions,
\begin{eqnarray}
\cos D_{\aster} & = & \sin \delta_0 \sin \delta_{\aster} + \cos \delta_0 \cos
 \delta_{\aster} \cos \Delta \alpha_{\aster}, \nonumber \\[0.01cm]
\cot \Pi_{\aster} & = & \cos \delta_0 \tan \delta_{\aster} \csc \Delta
 \alpha_{\aster} - \sin \delta_0 \cot \Delta \alpha_{\aster},
\end{eqnarray}
where \mbox{$\Delta \alpha_{\aster} = \alpha_{\aster} \!-\!  \alpha_0$}.  To
convert the particle's position from the cometocentric coordinate system to
the coordinate system centered on the center of a plate, we first compute right
ascension $\alpha_{\aster}$ and declination $\delta_{\aster}$ of the particle
from the formulas
\begin{equation}
\tan \Delta \alpha_{\aster} = \frac{\sin^2 \!\! D_{\aster} \tan
\Pi_{\aster} \! \left(\sin \delta_0 \!+ \cot \! D_{\aster} \cos \delta_0
 \sec \Pi_{\aster} \right)}{\cos^2 \!\! D_{\aster}(1 \!+ \cos^2 \! \delta_0
 \tan^2 \! \Pi_{\aster}) \!- \sin^2 \! \delta_0}, 
\end{equation}
and
\begin{equation}
\tan \delta_{\aster} = \cos \Delta \alpha_{\aster} \! \tan \delta_0
(1 \!+\! \cot \Pi_{\aster} \! \tan \Delta \alpha_{\aster} \csc \delta_0).
\end{equation}

\noindent
Since the sign of $\cos \Delta \alpha_{\aster}$ has a direct effect on the
sign of $\tan \delta_{\aster}$ and therefore also on the signs of $\sin
\delta_{\aster}$ and $\cos D_{\aster}$, the quadrant of $\Delta
\alpha_{\aster}$ in Equation (20) needs to be chosen such that after
inserting the values of $\Delta \alpha_{\aster}$ and $\delta_{\aster}$
from (21) into the first equation of (19) one gets the correct value of
$D_{\aster}$ and not its supplement.

Next, identifying $\alpha_0$ and $\delta_0$ with the coordinates of the plate
center, rather than the comet's nucleus, and inserting them together with the
particle's coordinates $\alpha_{\aster}$ and $\delta_{\aster}$ into Equations
(19), one obtains an angular distance $D_{\aster}$ and a position angle
$\Pi_{\aster}$ that determine the particle's position relative to the plate
center.\\[-0.4cm]

\begin{center}
{\footnotesize REFERENCES}
\end{center}
\vspace*{-0.5cm}
\begin{description}
{\footnotesize
\item[\hspace{-0.3cm}]
Brueckner, G. E., Howard, R. A., Koomen, M. J., et al. 1995, Sol.
{\hspace*{-0.6cm}}Phys., 162, 357
\\[-0.57cm]
\item[\hspace{-0.3cm}]
Cooper, T. P. 2003, MNASSA, 62, 170
\\[-0.57cm]
\item[\hspace{-0.3cm}]
Cooper, T. P. 2005, MNASSA, 64, 118
\\[-0.57cm]
\item[\hspace{-0.3cm}]
Cunningham, L. E. 1946a, HAC, 733
\\[-0.57cm]
\item[\hspace{-0.3cm}]
Cunningham, L. E. 1946b, IAUC, 1025
\\[-0.57cm]
\item[\hspace{-0.3cm}]
Gould, B. A. 1883, Astron. Nachr., 104, 129
\\[-0.57cm]
\item[\hspace{-0.3cm}]
Green, D. W. E. 2007, IAUC, 8883
\\[-0.57cm]
\item[\hspace{-0.3cm}]
Grindlay, J., Tang, S., Los, E., \& Servillat, M. 2012, IAU Symp.,{\linebreak}
{\hspace*{-0.6cm}}285, 29
\\[-0.57cm]
\item[\hspace{-0.3cm}]
Haddelsey, S. 2014, Operation Tabarin: Britain's Secret Wartime\linebreak
{\hspace*{-0.6cm}}Expedition to Antarctica 1944--1946. History Press,
Stroud, UK
\\[-0.57cm]
\item[\hspace{-0.3cm}]
Hockey, T. 2009, BAAS, 41, 572
\\[-0.57cm]
\item[\hspace{-0.3cm}]
Howard, R. A., Moses, J. D., Vourlidas, A., et al. 2008, Space Sci.
{\hspace*{-0.6cm}}Rev., 136, 67
\\[-0.07cm]
\item[\hspace{-0.3cm}]
Knight, M. M., A'Hearn, M. F., Biesecker, D. A., et al. 2010, AJ,{\linebreak}
{\hspace*{-0.6cm}}139, 926
\\[-0.57cm]
\item[\hspace{-0.3cm}]
Kreutz, H. 1888, Publ. Sternw. Kiel, 3
\\[-0.57cm]
\item[\hspace{-0.3cm}]
Kreutz, H. 1891, Publ. Sternw. Kiel, 6
\\[-0.57cm]
\item[\hspace{-0.3cm}]
Kreutz, H. 1901, Astron. Abh., 1, 1
\\[-0.57cm]
\item[\hspace{-0.3cm}]
Marsden, B. G. 1967, AJ, 72, 1170
\\[-0.57cm]
\item[\hspace{-0.3cm}]
Marsden, B. G. 1989, AJ, 98, 2306
\\[-0.57cm]
%
%
\item[\hspace{-0.3cm}]
Marsden, B. G. 2008, MPEC 2008-G44
\\[-0.57cm]
\item[\hspace{-0.3cm}]
Marsden, B. G., \& Williams, G. V. 2008, Catalogue of Cometary
{\hspace*{-0.6cm}}Orbits 2008, p.\ 108 (17th ed.; Cambridge, MA:
Smithsonian {\hspace*{-0.6cm}}Astrophysical Observatory, 195pp)
\\[-0.57cm]
\item[\hspace{-0.3cm}]
Marsden, B.\,G., Sekanina, Z., \& Yeomans, D.\,K.\ 1973, AJ,\,78,\,211
\\[-0.57cm]
\item[\hspace{-0.3cm}]
Michels, D. J., Sheeley, N. R., Howard, R. A., \& Koomen, M. J.{\linebreak}
{\hspace*{-0.6cm}}1982, Science, 215, 1097
\\[-0.57cm]
\item[\hspace{-0.3cm}]
Nagahara, H., Mysen, B. O., \& Kushiro, I. 1994, Geochim.
{\hspace*{-0.6cm}}Cosmochim. Acta, 58, 1951
\\[-0.57cm]
\item[\hspace{-0.3cm}]
Paraskevopoulos, J. S. 1945, IAUC, 1024
\\[-0.57cm]
\item[\hspace{-0.3cm}]
Pasachoff, J. M., Ru\v{s}\'{\i}n, V., Druckm\"{u}ller, M., et al.\ 2009,
ApJ, 702,{\linebreak} {\hspace*{-0.6cm}}1297
\\[-0.57cm]
\item[\hspace{-0.3cm}]
Sekanina, Z. 1978, QJRAS, 19, 52, 57
\\[-0.57cm]
%
%
\item[\hspace{-0.3cm}]
Sekanina, Z. 1982, AJ, 87, 1059
\\[-0.57cm]
\item[\hspace{-0.3cm}]
Sekanina, Z. 1984, Icarus, 58, 81
\\[-0.57cm]
\item[\hspace{-0.3cm}]
Sekanina, Z. 2000, ApJ, 545, L69
\\[-0.57cm]
\item[\hspace{-0.3cm}]
Sekanina, Z. 2002, ApJ, 566, 577
\\[-0.57cm]
\item[\hspace{-0.3cm}]
Sekanina, Z., \& Chodas, P. W. 2002, ApJ, 581, 760
\\[-0.57cm]
\item[\hspace{-0.3cm}]
Sekanina, Z., \& Chodas, P. W. 2004, ApJ, 607, 620
\\[-0.57cm]
\item[\hspace{-0.3cm}]
Sekanina, Z., \& Chodas, P. W. 2007, ApJ, 663, 657
\\[-0.57cm]
%
%
\item[\hspace{-0.3cm}]
Sekanina, Z., \& Chodas, P. W. 2012, ApJ, 757, 127 (33pp)
\\[-0.57cm]
\item[\hspace{-0.3cm}]
Sekanina, Z., \& Kracht, R. 2013, ApJ, 778, 24 (13pp)
\\[-0.57cm]
\item[\hspace{-0.3cm}]
Sekanina, Z., \& Kracht, R. 2015, ApJ, 801, 135 (19pp)
\\[-0.57cm]
\item[\hspace{-0.3cm}]
Simcoe, R. J., Grindlay, J. E., Los, E. J., et al. 2006, in~Applica-
{\hspace*{-0.6cm}}tions~of Digital Image Processing
XXIX,\,ed.\,A.\,G.\,Tescher,\,Proc.
{\hspace*{-0.6cm}}SPIE,~vol. 6312, 631217 (Bellington, WA).
\\[-0.57cm] 
\item[\hspace{-0.3cm}]
Thompson, W. T. 2009, Icarus, 200, 351
\\[-0.57cm]
\item[\hspace{-0.3cm}]
Tsvetkov, M., \& Tsvetkova, K. 2012, IAU Symp., 285, 417
\\[-0.57cm]

\item[\hspace{-0.3cm}]
van Heerden, H. J. 2008, MNASSA, 67, 116
\\[-0.65cm]
\item[\hspace{-0.3cm}]
Ye, Q.-Z., Hui, M.-T., Kracht, R., \& Wiegert, P. A. 2014,{\vspace{-0.05cm}}
 ApJ,{\linebreak}
 {\hspace*{-0.6cm}}796, 83 (8pp)}
\\[-0.57cm]
\vspace*{-0.46cm}
\end{description}
\end{document}